\documentclass{jfm}
\usepackage{graphicx,amssymb}
\usepackage{epstopdf, epsfig}

\shorttitle{Plane turbulent wall jets}
\shortauthor{Abhishek Gupta et al.}

\title{Mean velocity scaling in plane turbulent wall jets}

\author{Abhishek~Gupta\aff{1,2}, Harish~Choudhary\aff{1}, A.~K.~Singh\aff{2}, Thara~Prabhakaran\aff{1}
 \and Shivsai~Ajit~Dixit\aff{1}}

\affiliation{\aff{1}Indian Institute of Tropical Meteorology, Pashan, Pune 411008, India
\aff{2}Department of Physics, Institute of Science, Banaras Hindu University, Varanasi 221005, India}

\begin{document}

\maketitle

\begin{abstract}
Studies in the literature on plane turbulent wall jets on flat surfaces, have invariably considered either the nozzle initial conditions or the asymptotic conditions far downstream, as scaling parameters for the streamwise variations of length and velocity scales. These choices, however, do not square with the notion of self similarity which is essentially a ``local" concept. We first demonstrate that the streamwise variations of velocity and length scales in wall jets show remarkable scaling with local parameters i.e. there appear to be no imposed length and velocity scales. Next, it is shown that the mean velocity profile data suggest existence of two distinct layers - the wall (inner) layer and the full-free jet (outer) layer. Each of these layers scales on the appropriate length and velocity scales and this scaling is observed to be universal i.e. independent of the local friction Reynolds number. Analysis shows that the overlap of these universal scalings leads to a Reynolds-number-dependent power-law velocity variation in the overlap layer. It is observed that the mean-velocity overlap layer corresponds well to the momentum-balance mesolayer and there appears to be no evidence for an inertial overlap; only the meso-overlap is observed. Introduction of an intermediate variable absorbs the Reynolds-number dependence of the length scale in the overlap layer and this leads to a universal power-law overlap profile for mean velocity in terms of the intermediate variable.
\end{abstract}


\section{Introduction}\label{sec:intro}
Starting with the work of \cite{glauert1956} in the mid-fifties, turbulent wall jets developing on flat surfaces have kept researchers engaged for quite some time. The studies range from experiments \citep{schwarz1961,bradshaw1962,myers1963,tailland1967,irwin1973,wygnanski1992,schneider1994,
eriksson1998,tachie2002,rostamy2011,tang2015,gnana2019} to theoretical approaches \citep{glauert1956,irwin1973,narasimha1973,george2000,afzal2005,barenblatt2005,gersten2015} with some classic intermediate reviews \citep{launder1979,launder1983} on the subject. With the increase in computing power, direct numerical simulations \citep[DNS,][]{ahlman2007,naqavi2018} and large-eddy simulations \citep[LES,][]{dejoan2005,banyassady2014,banyassady2015} at moderate Reynolds numbers have become possible in the recent times. Wall jets, unlike other canonical wall-bounded turbulent flows, are characterized by a non-monotone mean velocity profile with a velocity maximum occurring close to the surface. On either side of this maximum, velocity decreases in the wall normal direction (see figure~\ref{fig:setup}\textit{a}). It is also known that a region of counter-gradient momentum diffusion occurs below the velocity maximum in wall jets \citep{narasimha1990}. Notwithstanding such unique features, advances in the understanding of wall jets have remained rather sporadic possibly due to fewer engineering applications - the two most prominent examples being (a) slot blowing used for separation control on the suction side of an aerofoil and (b) coolant flows in electronic devices and turbine blades \citep{launder1983}. 

Scaling mean velocity in plane (two-dimensional), fully-developed, turbulent wall jets (henceforth wall jets) has proved to be challenging and typically involves scaling: (i) streamwise variations of the velocity and length scales and (ii) velocity profiles $U(z)$ using these scales. In this work, $x$, $y$ and $z$ respectively denote streamwise, spanwise and wall-normal coordinates. Experiments indicate that wall jets can be considered fully-developed beyond $x/b\approx30$ \citep[][henceforth NYP and EKP respectively]{narasimha1973,eriksson1998}. Note that all mean integral quantities such as length scales, velocity scales etc. in a fully-developed wall jet continue to ``develop" (vary) in the $x$ direction. This is to be contrasted with fully-developed \emph{internal} flows (pipes and channels) wherein mean-flow development is absent.

For a wall jet flow, the two important velocity scales of interest are the maximum velocity $U_{max}$ (figure~\ref{fig:setup}\textit{a}) and, the friction velocity $U_{\tau}=\sqrt{\tau_{w}/\rho}$; $\tau_{w}$ is the wall shear stress and $\rho$ is the fluid density. Similarly, the height $z_{T}$ from the wall above the velocity maximum where velocity equals $U_{max}/2$ (figure~\ref{fig:setup}\textit{a}) and the viscous length $\nu/U_{\tau}$ are the two important length scales; $\nu$ is the fluid kinematic viscosity and $z_{T}$ represents overall thickness of the flow. The ratio of these length scales is the local Reynolds number $\Rey_{\tau}=z_{T}U_{\tau}/\nu$. The choice of $z_{T}$ in lieu of $z_{max}$ (see NYP) is motivated by the fact that the shape of the velocity profile is rather flat in the region around the velocity maximum. This makes $z_{max}$ more prone to measurement errors leading to larger data scatter that hampers testing of scaling laws. On the other hand, $z_{T}$ does not suffer from this drawback and is usually quite accurately obtained in experiments. Nozzle parameters such as the exit velocity $U_{j}$ (in the potential core), slot height $b$, Reynolds number $Re_{j}=U_{j}b/\nu$ and kinematic momentum rate per unit width (also sometimes referred to as the momentum flux) $M_{j}\sim U_{j}^{2}b$ constitute a set of \emph{initial conditions} (ICs) for the wall jet flow.   

In the case of streamwise development of length and velocity scales $U_{max}$, $z_{T}$ and $U_{\tau}$ (see point (i) above), most scaling approaches till date \citep{glauert1956,narasimha1973,george2000,barenblatt2005} have used, in some form, one or more nozzle ICs as scaling parameters. Alternatively in a more recent theoretical work, \cite{gersten2015} has noted that, in certain (but not all) experimental configurations, the wall jet flow could asymptotically tend to a half-free jet as Reynolds number tends to infinity (to be discussed later in some detail). Flow properties in this asymptotic state may be termed as the \emph{far downstream conditions} (FCs); \cite{gersten2015} has proposed the momentum rate ($M_{\infty}$) of the asymptotic half-free jet as the relevant scaling parameter. These considerations lead to a fundamental question: Does a wall jet always ``remember" the ICs or ``know" the FCs throughout its development? An affirmative answer implies that truly self-similar development, controlled essentially by the local parameters, is not possible. 

Possible influence of ICs(FCs) on the scaling of mean velocity profile (see point (ii) above) can be briefly outlined as follows. In general, the structure of spatially developing turbulent wall-bounded flows consists of different layers, that follow different \emph{local} (localized in the $x$ direction) scalings, i.e. layer-wise self-similarity, but develop downstream at different rates. Such situations typically result in downstream increase of $\Rey_{\tau}$, as for example in turbulent boundary layers (TBLs). In wall jets, there could be local $\Rey_{\tau}$-dependence (non-universality) of the inner and outer mean velocity scaling laws and hence their overlap \citep[as in][]{george2000}. The dependence of the length and velocity scales on ICs(FCs), if any, then presents an additional but distinct complication that could render the velocity profile scaling in wall jets further dependent on the ICs \citep{barenblatt2005} or FCs \citep{gersten2015} in addition to the local $\Rey_{\tau}$-dependence. 

In this work, we use data from our own experiments and the literature (\S~\ref{sec:expt} and \S~\ref{sec:localscaling}), and demonstrate that the streamwise variations of $U_{max}$, $z_{T}$ and $U_{\tau}$ better scale with the ``local" kinematic momentum rate $M=\int_{0}^{\infty}U^{2}\textrm{d}z$ (per unit width) and $\nu$ i.e. \emph{local conditions} (LCs) than ICs and FCs. In other words, there appears to be no dependence on ICs and FCs, and the basic requirement for self similarity, namely the lack of imposed length and velocity scales, appears to be satisfied (\S~\ref{sec:localscaling}). Next, it is shown that the wall jet flow comprises of universal (independent of both, nozzle ICs/FCs and $\Rey_{\tau}$) inner and outer scaling layers (\S~\ref{subsec:innerouterscaling}) termed as the wall layer and the full-free jet layer respectively. Starting with this two-layer structure, theoretical arguments, free from any modelling assumptions, show that the overlap of these self-similar (universal) layers leads to a non-universal ($\Rey_{\tau}$-dependent) power-law profile for mean velocity in the overlap layer (\S~\ref{subsec:overlapanalysis} and \S~\ref{subsec:overlapexptdata}). Analysis of DNS data reveals that the mean-velocity overlap layer corresponds quite well with the momentum-balance mesolayer in wall jets (\S~\ref{subsec:layers}) and this correspondence may be used to effectively absorb the $\Rey_{\tau}$-dependence of the overlap layer into an intermediate variable which in turn leads to a universal power-law mean velocity profile in the overlap layer. Data also indicate that there is no inertial overlap between the inner and outer scaling regions for the range of Reynolds numbers considered here (\S~\ref{subsec:layers}).   

\section{Experimental details}\label{sec:expt}

\begin{figure*}
\centerline{\includegraphics[width=0.9\textwidth]{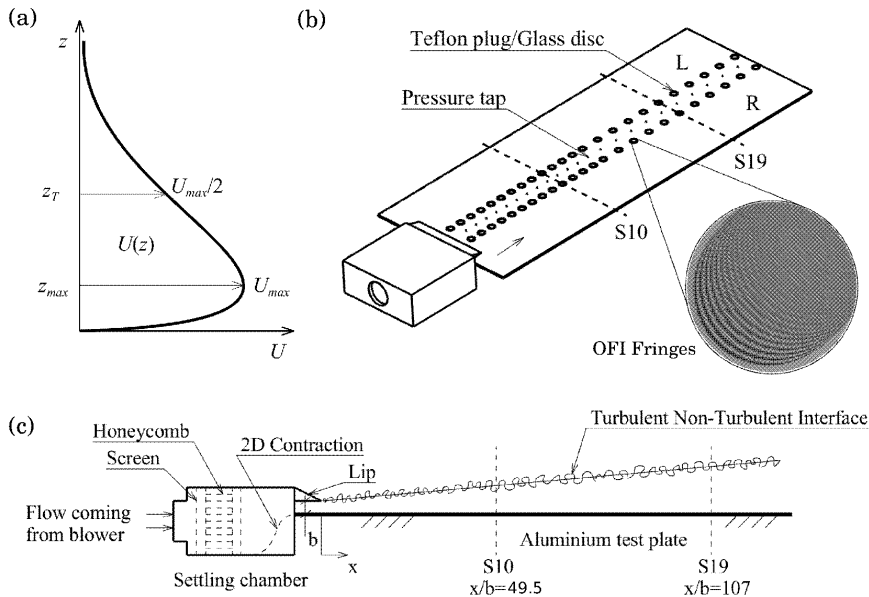}}
  \caption{(\textit{a}) Typical mean velocity profile in a turbulent wall jet with essential definitions. Schematic (\textit{b}) isometric and (\textit{c}) side views of the wall jet setup at the Fluid Dynamics Laboratory (FDL), IITM, Pune. Region of interest (ROI) is from stations S10 to S19 where fully-developed and nominally two-dimensional mean flow is obtained. Zoomed view of the sample OFI fringes obtained on an SF11 glass disc in the ROI is also shown in (\textit{b}). $L$ and $R$ respectively denote OFI locations on the left and right sides of the test-surface centerline.}
\label{fig:setup}
\end{figure*}

\begin{figure}
  \centerline{\includegraphics[width=0.9\textwidth]{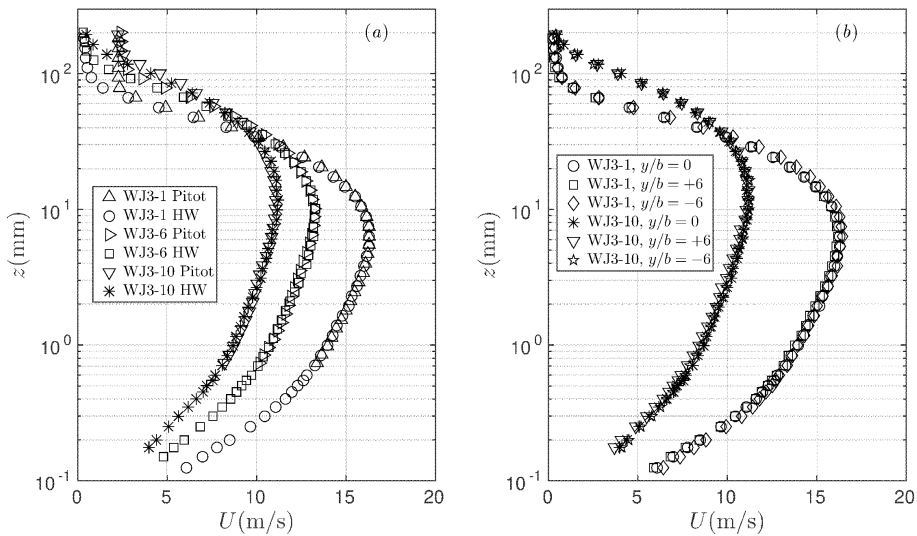}}%
  \caption{Mean velocity data for $Re_{j}=21228$. (\textit{a}) Dimensional velocity profiles measured by Pitot tube and hotwire (HW) probe at different streamwise stations and (\textit{b}) dimensional HW profiles at three spanwise stations and two streamwise stations.}
\label{fig:twodimselfsim}
\end{figure}

Figures~\ref{fig:setup}(\textit{b}) and \ref{fig:setup}(\textit{c}) show the wall jet setup constructed at the Fluid Dynamics Laboratory (FDL), Indian Institute of Tropical Meteorology (IITM), Pune, India. A settling chamber, consisting of a honeycomb and a set of suitable screens, admits air from a well-balanced, high-flow rate blower (Ametek Nautilair $8.9''$ impeller diameter) and discharges it through a two-dimensional nozzle (width $L=300$ mm, height $b=10$ mm) tangentially onto the test surface. Blower speed may be adjusted using a PWM controller and kept constant even under variable load conditions. For well-defined initial conditions, a sharp aluminium upper lip is fitted at the exit of the nozzle. A recent study \citep{mcintyre2019} focuses on the effect of thickness of a blunt lip on the downstream development of wall jets and concludes that no significant effects of the lip thickness are observed on the profiles of various quantities beyond $x/b\approx10$. However, McIntyre \emph{et al.} have not discussed the effect of lip thickness on spanwise homogenity of the mean flow. In our setup, we have observed that use of a blunt upper lip results in significant spanwise variations of mean flow quantities even at distances far downstream from the nozzle. However, it is observed, that a fair stretch of nominally two-dimensional flow (figure~\ref{fig:twodimselfsim}\textit{b}) is obtained when one resorts to a sharp upper lip instead of a blunt one. The lower lip of the nozzle is carefully leveled with the test surface. Velocity profiles at nozzle exit are close to the top-hat profile (not shown) and uniform across the entire width except for small portions near the ends. The test surface is a flat, straight and polished aluminium plate (width $600$ mm, length $1500$ mm and thickness $6$ mm). The size of the room is large enough for secondary flow effects to be minimal. Static pressure ports of $0.5$ mm diameter are drilled in the brass bushes fitted along the longitudinal centerline of the test surface. Counterbore brass bushes are fitted on either side (\emph{L} and \emph{R} in figure~\ref{fig:setup}\textit{b}) of the longitudinal centerline and are used to mount the SF11 glass discs for Oil Film Interferometry (OFI); unused bushes are fitted with Teflon plugs. It is ensured that all fittings are flush with the test surface. 

Mean velocity profiles at different streamwise and spanwise stations are measured, using a Pitot tube and a single hotwire probe, for three nozzle Reynolds numbers $Re_{j}=U_{j}b/\nu=10244$, $15742$ and $21228$ (WJ1, WJ2 and WJ3 respectively in table~\ref{tab:data}); $\nu=1.5\times10^{-5}\mathrm{m}^{2}\mathrm{s}^{-1}$ is the kinematic viscosity of air at room temperature prevalent during experiments. Pitot tube has an outer diameter of $1.2$~mm and an ethanol-based projection manometer having least count of $0.1$~mm of ethanol column is used to measure the pressure difference between the Pitot-tube and a reference static-pressure port on the test surface. It is verified that there is no gradient of mean static pressure in the $x$ direction. Near-wall Pitot readings are corrected using the procedure given by \cite{bailey2013}. The single hotwire probe is custom-made with prongs made of sharp stainless steel needles with the tip spacing of about $2.5$ mm. The sensor is a silver cladded Pt-Rh Wollaston wire of core diameter $d=5$ $\mathrm{\mu m}$. The wire is first soldered to the prongs and silver cladding in the central portion is etched away electrochemically using $10\%$ nitric acid solution to expose the Pt-Rh sensor element. The active length is $l\approx 0.9$ mm so that $l/d\approx 180$ and $19\leq l_{+}\leq 56$ over the entire range of experiments; $l_{+}=l U_{\tau}/\nu$ is the length of the sensor in wall units. The sensor is operated by the StreamLine Pro constant temperature anemometer from Dantec Dynamics, Denmark and the anemometer output is acquired at $10$ kHz on a computer through National Instruments PCI-6143 card using Dantec's StreamWare Pro software. Calibration of the sensor is performed \emph{in situ} by recording the mean anemometer voltage $V$ (sampled at $10$ kHz for $30$ seconds) over a range of blower speeds, at a fixed height from the wall (near the velocity maximum); mean velocity $U$ at this height is measured by the Pitot tube. King's law calibration equation of the form $V^{2}=\zeta_{1}+\zeta_{2} U^{0.45}$ is used to obtain calibration constants $\zeta_{1}$ and $\zeta_{2}$. Calibration is performed before and after each experiment and data with significant calibration drifts are discarded. It is ensured that the room temperature variations during the experiments are within $\pm 0.5^{\circ}\textrm{C}$ of the average temperature during the experiment. Data during the experiments are sampled at $10$ kHz for $60$, $90$ and $120$ seconds respectively for $Re_{j}=21228$, $15742$ and $10244$ in order to compensate for the lower overall flow speeds at lower Reynolds numbers. Figure~\ref{fig:twodimselfsim}(\textit{a}) shows Pitot tube and hotwire measurements agreeing quite well with each other demonstrating measurement consistency. Pitot readings saturate to constant non-zero values away from the wall due to the insensitivity of the Pitot-tube-alcohol manometer system to small velocity variations. Hotwire data, on the other hand, correctly show the velocities tending to zero far away from the wall. Figure~\ref{fig:twodimselfsim}(\textit{b}) shows collapse of dimensional hotwire profiles demonstrating two-dimensionality of the mean flow over the streamwise-spanwise extent $49.5\leq x/b \leq 107$ and $-6\leq y/b \leq 6$.

Wall shear stress $\tau_{w}$ is directly measured at $y/b\approx\pm4.5$ on either side of the plate centerline (see $L$ and $R$ in figure~\ref{fig:setup}\textit{b}) using OFI technique. A small drop of silicone fluid is placed on an SF11 glass disc (blackened on its bottom side) fitted flush in the brass bush. Smearing of the drop due to the flow forms a film that thins down at a rate proportional to $\tau_{w}$. When illuminated by a sodium vapour lamp (average $\lambda\approx589.3$ $\mathrm{nm}$), the reflected near-monochromatic light from the top and bottom interfaces of the film forms an interference pattern (figure~\ref{fig:setup}\textit{b}); $\tau_{w}$ is related to the time rate of increase $\mathrm{\Delta} x/\mathrm{\Delta} t$ of the inter-fringe spacing \citep{chauhan2010}. A computer-controlled DSLR camera (Nikon D5500) fitted with a prime macro lens (AF-S VR Micro-Nikkor 105mm f/2.8G IF-ED) is used to capture the interference pattern at every $2$ seconds interval. The camera is located well above the plate at a viewing angle $\theta$ which in turn is measured from a photograph of the setup taken using another camera. Image sequences with significant dust contamination of the oil film are discarded. Each image is converted to gray scale and subjected to FFT analysis of pixel intensity using custom-written MATLAB codes. A robust estimate of the fringe spacing $\mathrm{\Delta} x$ in each image is extracted by averaging over at least $50$ pixel rows in the region where the fringes are straight and perpendicular to the overall flow direction. In order to ensure consistency and repeatability, silicone fluids of two different nominal viscosities ($\nu_{s}=100$ and $200$ cSt) are used for all measurements. These fluids are calibrated for their density $\rho_{s}$, refractive index $n_{s}$ and temperature dependence of kinematic viscosity $\nu_{s}$. Temperature of the thin film, just before and after each experiment, is measured with a hand-held infrared thermometer (Fluke 64 MAX); this allows appropriate calculation of $\nu_{s}$ under the experimental conditions. Relation (1) from \cite{chauhan2010} is used to compute $\tau_{w}$ and $U_{\tau}$. In each experimental run, values of $U_{\tau}$ from both sides of the centerline and with both the silicone fluids are found to agree to within $\pm2.5\%$ of the average of those four values; this average value is given in table~\ref{tab:data}. This reconfirms mean flow two-dimensionality. Table~\ref{tab:data} lists important experimental parameters for all our data sets (WJ1, WJ2 and WJ3) used in this work.

\begin{table}
  \begin{center}
\def~{\hphantom{0}}
  \begin{tabular}{lcccllccccc}
      Flow & $U_{j}$ & $b$ & $Re_{j}$ & Data & $x/b$ & $U_{max}$ & $U_{\tau}$ & $\Rey_{\tau}$ & $M$ & $M_{\infty}$\\
 Code & (ms$^{-1}$) & (mm) & & Code & & (ms$^{-1}$) & (ms$^{-1}$) & & (m$^{3}$s$^{-2}$) & (m$^{3}$s$^{-2}$) \\[2pt]
WJ1  & 15.37 & 10 & 10244 & WJ1-1 & 49.5 & 7.60 & 0.4860 & 1314 & 1.733 & 0.719\\
          &  & & & WJ1-2 & 54.5 & 7.27 & 0.4618 & 1363 & 1.745 & 0.719\\
          &  & & & WJ1-3 & 59.5 & 6.89 & 0.4395 & 1408 & 1.679 & 0.717\\
          &  & & & WJ1-4 & 64.5 & 6.74 & 0.4222 & 1452 & 1.713 & 0.719\\
          &  & & & WJ1-5 & 69.5 & 6.53 & 0.4038 & 1511 & 1.753 & 0.728\\
          &  & & & WJ1-6 & 77 & 6.22 & 0.3855 & 1555 & 1.717 & 0.723\\
          &  & & & WJ1-7 & 84.5 & 5.93 & 0.3712 & 1661 & 1.745 & 0.764\\
          &  & & & WJ1-8 & 92 & 5.62 & 0.3493 & 1678 & 1.674 & 0.723\\
          &  & & & WJ1-9 & 99.5 & 5.50 & 0.3312 & 1644 & 1.658 & 0.672\\
          &  & & & WJ1-10 & 107 & 5.24 & 0.3224 & 1733 & 1.636 & 0.705\\[2pt]

WJ2  & 23.61 & 10 & 15742 & WJ2-1 & 49.5 & 11.89 & 0.7230 & 1880 & 4.062 & 1.770\\
          &  & & & WJ2-2 & 54.5 & 11.37 & 0.6884 & 1962 & 4.099 & 1.788\\
          &  & & & WJ2-3 & 59.5 & 10.90 & 0.6569 & 2058 & 4.120 & 1.824\\
          &  & & & WJ2-4 & 64.5 & 10.44 & 0.6272 & 2104 & 4.046 & 1.795\\
          &  & & & WJ2-5 & 69.5 & 10.25 & 0.6005 & 2189 & 4.232 & 1.816\\
          &  & & & WJ2-6 & 77 & 9.63 & 0.5706 & 2239 & 4.034 & 1.780\\
          &  & & & WJ2-7 & 84.5 & 9.30 & 0.5456 & 2338 & 4.083 & 1.806\\
          &  & & & WJ2-8 & 92 & 8.70 & 0.5155 & 2410 & 3.905 & 1.780\\
          &  & & & WJ2-9 & 99.5 & 8.52 & 0.4928 & 2428 & 3.950 & 1.719\\
          &  & & & WJ2-10 & 107 & 8.08 & 0.4713 & 2503 & 3.838 & 1.714\\[2pt]

WJ3  & 31.84 & 10 & 21228 & WJ3-1 & 49.5 & 16.29 & 0.9373 & 2557 & 8.029 & 3.510\\
          &  & & & WJ3-2 & 54.5 & 15.52 & 0.8968 & 2692 & 8.058 & 3.605\\
          &  & & & WJ3-3 & 59.5 & 14.92 & 0.8557 & 2801 & 8.065 & 3.630\\
          &  & & & WJ3-4 & 64.5 & 14.37 & 0.8158 & 2821 & 7.927 & 3.496\\
          &  & & & WJ3-5 & 69.5 & 13.80 & 0.7868 & 2969 & 7.974 & 3.615\\
          &  & & & WJ3-6 & 77 & 13.30 & 0.7423 & 3023 & 7.950 & 3.495\\
          &  & & & WJ3-7 & 84.5 & 12.54 & 0.7079 & 3195 & 7.860 & 3.595\\
          &  & & & WJ3-8 & 92 & 12.08 & 0.6685 & 3222 & 7.774 & 3.433\\
          &  & & & WJ3-9 & 99.5 & 11.55 & 0.6380 & 3259 & 7.581 & 3.328\\
          &  & & & WJ3-10 & 107 & 11.20 & 0.6081 & 3274 & 7.511 & 3.191\\[2pt]
          
EKP1 & 1.04 & 9.6 & 9600 & EKP1-1 & 40 & 0.57 & 0.0340 & 1140 & 0.008 & 0.004\\
          &  & & & EKP1-2 & 70 & 0.42 & 0.0254 & 1394 & 0.008 & 0.004\\
          &  & & & EKP1-3 & 100 & 0.34 & 0.0202 & 1621 & 0.007 & 0.004\\
          &  & & & EKP1-4 & 150 & 0.27 & 0.0158 & 1757 & 0.007 & 0.003\\[2pt]
          
SC1   & 25.30 & 10 & 42839 & SC1-1 & 30 & 17.35 & - & - & 13.224 & -\\
          &  & & & SC1-2 & 36 & 15.91 & - & - & 13.367 & -\\
          &  & & & SC1-3 & 42 & 14.55 & - & - & 12.400 & -\\[2pt]
          
BG1 & 198 & 0.46 & 6113 & BG1-1 & 585 & 24.04 & 1.4207 & 2087 & 9.390 & 4.643\\
          &  & & & BG1-2 & 793 & 20.46 & 1.1882 & 2294 & 8.962 & 4.501\\
          &  & & & BG1-3 & 999 & 18.21 & 1.0436 & 2530 & 8.928 & 4.559\\
          &  & & & BG1-4 & 1225 & 16.36 & 0.9294 & 2703 & 8.651 & 4.467\\[2pt]
          
TM1 & 26.77 & 6 & 11000 & TM1-1 & 133 & 8.92 & 0.4919 & 2450 & 3.830 & 1.831\\
TM2 & 43.2 & 6 & 18000 & TM2-1 & 133 & 14.95 & 0.7859 & 3742 & 10.416 & 5.108\\
TM3 & 60 & 6 & 25000 & TM3-1 & 100 & 25.36 & 1.3101 & 4469 & 21.295 & 10.788\\
          &  & & & TM3-2 & 167 & 19.51 & 0.9789 & 5472 & 20.555 & 10.539\\[2pt]
          
NTT1 & 11.25 & 10 & 7500 & NTT1-1 & 35 & ~7.65 & 0.4648 & 865 & ~1.227 & -\\
  \end{tabular}
  \caption{Important parameters for the data sets under consideration. WJ1, WJ2 and WJ3 are the present experimental data sets. EKP1 - \cite{eriksson1998}, SC1 - \cite{schwarz1961}, BG1 - \cite{bradshaw1962} and TM1, TM2 and TM3 - \cite{tailland1967} are experimental data sets from the literature. NTT1 - \cite{naqavi2018} is the recent DNS data set. $U_{\tau}$ values for BG and TM data sets have been corrected as discussed in \textsection\ref{subsec:data}. $M_{\infty}$ is computed as per the procedure given in \textsection 8 of \cite{gersten2015}. $M_{\infty}$ is not computed for SC data due to unavailability of $U_{\tau}$ and NTT data due to only a single data point.}
  \label{tab:data}
  \end{center}
\end{table}

\section{Scaling streamwise variations of $U_{max}$, $z_{T}$ and $U_{\tau}$}\label{sec:localscaling}

First, we shall give a brief account of the existing scaling approaches based on ICs and FCs in addition to the present proposal based on LCs. To remind the reader, we have three kinematic momentum rates (per unit width) of significance for ICs, FCs and LCs (see \textsection~\ref{sec:intro}) and they are: the nozzle momentum rate $M_{j}=U_{j}^{2}b$, the far-downstream momentum rate $M_{\infty}$ of the asymptotic half-free jet state and the local wall jet momentum rate $M=\int_{0}^{\infty}U^{2}\textrm{d}z$ respectively. Data sets selected from the literature will also be discussed briefly and the evaluation of various scaling approaches will be presented. 

\subsection{Approaches based on nozzle ICs}\label{subsec:ICs}

As noted by NYP, early approaches have relied upon $U_{j}$ and $b$ as the relevant scaling parameters i.e. relations of the form $U_{max}/U_{j}=F(x/b)$, where $F$ is a universal function, should hold. However, after compiling large amount of experimental data on wall jets in still air, NYP have found that data do not scale simply on $U_{j}$ and $b$. Based on the dynamical importance of the nozzle kinematic momentum rate $M_{j}$, NYP have proposed the relavant scaling parameters to be $M_{j}$ (instead of $U_{j}$ and $b$ separately) and $\nu$. NYP have showed that this $M_{j}$-$\nu$ scaling leads to better collapse of data with relationships of the form $U_{max}\nu/M_{j}=f_{1}\left(xM_{j}/\nu^{2}\right)$, $z_{T}M_{j}/\nu^{2}=f_{2}\left(xM_{j}/\nu^{2}\right)$ etc. where $f_{1}$ and $f_{2}$ are supposed to be universal functions. Several further studies \citep{wygnanski1992,george2000} have used NYP scaling to present their data and found it to be reasonably robust.

More recently, \cite{barenblatt2005} - henceforth BCP - have argued that the strong dependence on slot height $b$ remains important and this leads to the so-called incomplete similarity in the dimensionless independent variables $\Pi_{1}=z/b$, $\Pi_{2}=x/b$ and $\Pi_{3}=\sqrt{M_{j}b/\nu^{2}}$. Note that in the BCP framework, the governing (independent) parameters are taken to be $x$, $z$, $b$, $M_{j}$ and $\nu$, and $b$ and $\nu$ are used to make $x$, $z$ and $M_{j}$ dimensionless in accordance with the Buckingham's Pi theorem. Since $\Pi_{3}$ is a measure of the nozzle Reynolds number $Re_{j}$, streamwise variations of all the length and velocity scales may be expressed as $U_{max}b/\nu=f_{3}\left(x/b,\Rey_{j}\right)$, $z_{T}/b=f_{4}\left(x/b,\Rey_{j}\right)$ etc. For complete similarity according to BCP, this implies that the data from various experiments should collapse to universal curves in plots $U_{max}b/\nu$ versus $x/b$, $z_{T}/b$ versus $x/b$ etc. as $Re_{j}\rightarrow\infty$. However, the results of NYP indicate that $z_{T}/b$ versus $x/b$ plot does not show such collapse. This observation could be taken to support the incomplete similarity hypothesis of BCP.

As a side line, one may ask if there is any connection at all between these seemingly different approaches of NYP and BCP. To see this, consider choosing $M_{j}$ and $\nu$ (instead of $b$ and $\nu$ but still remaining within the BCP framework) to form the dimensionless groups. One then obtains $\Pi_{1}^{*}=zM_{j}/\nu^{2}$, $\Pi_{2}^{*}=xM_{j}/\nu^{2}$ and $\Pi_{3}^{*}=\Pi_{3}=\sqrt{M_{j}b/\nu^{2}}$. This choice leads to the scaling relations $U_{max}\nu/M_{j}=f_{3}^{*}\left(xM_{j}/\nu^{2},\Rey_{j}\right)$, $z_{T}M_{j}/\nu^{2}=f_{4}^{*}\left(xM_{j}/\nu^{2},\Rey_{j}\right)$ etc. Interestingly, NYP have already shown that the data (accessible to them) collapse fairly well to near-universal curves i.e. no $Re_{j}$-dependence, in plots $U_{max}\nu/M_{j}$ versus $xM_{j}/\nu^{2}$, $z_{T}M_{j}/\nu^{2}$ versus $xM_{j}/\nu^{2}$ etc. This suggests complete similarity! In other words, simple change of repeating variables in BCP approach leads to the approach of NYP which in turn supports complete similarity i.e. direct effect of slot height $b$, if any, appears to be negligibly small.

To summarize, the approach of NYP appears to be the most appropriate amongst those that use nozzle ICs as scaling parameters.

\subsection{Approach based on FCs}\label{subsec:FCs}

Recently, \cite{gersten2015} has pointed out the need for distinction between different wall jet configurations based on geometry of the experimental setup i.e. whether (or not) the nozzle is located in a large wall perpendicular to the test surface. This distinction is motivated by the theoretical results for turbulent plane \emph{free jets} by \cite{schneider1985} wherein it is shown that the kinematic momentum flux $M$ at large distances ($x\rightarrow\infty$) from the nozzle - (i) asymptotically vanishes ($M\rightarrow 0$) if the nozzle is located in a large plane wall perpendicular to the jet exit velocity vector and (ii) remains constant ($M=M_{j}$) for nozzles without any such wall. Extending this idea to \emph{wall jets}, \cite{gersten2015} has hypothesized that the asymptotic state far downstream (as Reynolds number tends to infinity) would be a half-free jet with finite momentum rate $M_{\infty}$ (we denote this by FC; see \textsection\ref{sec:intro}) if the nozzle in the wall jet setup is not located in a wall perpendicular to the test surface. With this, the wall jet flow could be considered as a perturbation of the limiting half-free jet state and therefore \cite{gersten2015} has proposed $M_{\infty}$ as the scaling parameter for the streamwise coordinate $x$. This leads to the definition of a local Reynolds number $\Rey_{x}=\sqrt{(x-x_{0})M_{\infty}}/\nu$ ($x_{0}$ is the adjustment for the virtual origin effect) and all the other dimensionless parameters are then expected to be universal functions of $\Rey_{x}$. Consistent with the current presentation, the expected universal relationships involving $M_{\infty}$ and $\nu$ may be written as $U_{max}\nu/M_{\infty}=f_{5}\left(xM_{\infty}/\nu^{2}\right)$, $z_{T}M_{\infty}/\nu^{2}=f_{6}\left(xM_{\infty}/\nu^{2}\right)$ etc.

\subsection{The present approach based on LCs}\label{subsec:LCs}

As outlined in \textsection\ref{sec:intro}, if indeed wall jet flow evolves downstream in a self-similar fashion, then the local kinematic momentum rate $M=\int_{0}^{\infty}U^{2}\textrm{d}z$ and $\nu$ (i.e. LCs) could very well be the candidate scaling parameters. Some support for this proposal may be derived from the work of NYP who have suggested ``A similar correlation using local momentum flux $M$ is implied by (3) $\ldots$ but clearly $M$ is less convenient than $M_{j}$". In case of self-similar development, data should show better collapse with LCs than with ICs or FCs. Therefore according to our present proposal of ``local" scaling, we expect universal relationships of the form $U_{max}\nu/M=f_{7}\left(xM/\nu^{2}\right)$, $z_{T}M/\nu^{2}=f_{8}\left(xM/\nu^{2}\right)$ etc. 

\subsection{Wall jet data sets from the literature}\label{subsec:data}

In order to evaluate the scaling approaches mentioned above, we have selected, apart from our own experimental data (\textsection{\ref{sec:expt}}), five well-documented studies from the literature. These are briefly described below with necessary parameters given in table~\ref{tab:data}.

The main constraints on the selection of the data sets are the availability of measured mean velocity profiles and accurate estimate of the wall shear stress $\tau_{w}$. The former is necessary to evaluate the local momentum rate $M$ and the latter is crucial for testing the scaling of skin friction. Accurate measurement of $\tau_{w}$ poses serious challenges in wall jets \citep{launder1979}. Computation of $\tau_{w}$ from the mean velocity gradient in the linear viscous sublayer is accurate only with non-heat transfer techniques such as Laser Doppler Velocimetry (LDV). Hotwire data are known to yield spurious large values of velocity in the near wall region due to strong conduction heat transfer to the wall \citep{durst2001}, the so-called wall proximity effect. This could result in reduction in the estimated values of velocity gradient at the wall leading to lower-than-actual values of $\tau_{w}$ (and hence $U_{\tau}$). Use of impact tube devices such as the Stanton tube relies on the universality of the mean velocity profile over the wall-normal extent of the device. Usually such devices are calibrated in a canonical flow such as a fully-developed channel or a zero-pressure-gradient TBL and then used in wall jets to measure skin friction \citep{bradshaw1962}. However, it has been reported that the linear velocity profile in the sublayer of wall jets extends only up to $z_{+}\approx 3$ in contrast to the celebrated $z_{+}\approx5$ in canonical TBLs, pipes and channels \citep{eriksson1998}. As we shall see later, such a reduction in the thickness of the linear sublayer is consistent with the strong influence of the outer jet flow on the inner wall flow.  In view of these difficulties, one may consider applying some judiciously estimated corrections to $\tau_{w}$ values measured in wall jets using hotwire mean velocity gradient and impact tubes. These corrections are discussed further in the relevant places. In our experiments described in \textsection{\ref{sec:expt}}, we have used OFI that does not suffer from these drawbacks and allows direct, reliable measurement of $\tau_{w}$.  

The first data set is from EKP and the data measured for $Re_{j}=9600$ at four different streamwise locations are selected for the present purpose. These are available at the ERCOFTAC Classic Collection Database. The working fluid is water and nozzle is located in a vertical wall perpendicular to the test surface. Mean velocity has been measured using high-resolution LDV and $\tau_{w}$ has been computed from velocity gradient in the linear viscous sublayer. There is reverse flow above the wall jet in this setup due to the vertical wall above the nozzle exit extending all the way to the free surface of water.   

The second set of data come from \cite{schwarz1961} - henceforth SC - and have been measured at different nozzle Reynolds numbers of which $Re_{j}=42839$ data have been digitized from figure~2 given in SC. The working fluid is air and nozzle is \emph{not} located in a wall perpendicular to the test surface. Mean velocity has been measured using hotwire anemometry and $\tau_{w}$ has not been measured in these experiments.

The third data set is from the extensive measurements by \cite{bradshaw1962} - hereafter BG. These are also air-based experiments with $Re_{j}=6113$ (see table~\ref{tab:data}) and four representative hotwire mean velocity profiles have been digitized from the figures~2, 3 and 6 in BG. Nozzle is \emph{not} located in a wall perpendicular to the test surface and $\tau_{w}$ has been measured using Stanton tube \citep{bradshaw1961,bradshaw1962} made typically from a $0.002$ inches ($z_{+}\approx3.5$ to $6$ for BG data) thick steel blade; the Stanton tube has been calibrated in a channel flow and used to measure $\tau_{w}$ in the wall jets. However as mentioned before, the linear sublayer extends only up to $z_{+}\approx3$ in wall jets (see figure~10 of EKP) as against $z_{+}\approx5$ in channel flows. Therefore, the Stanton tube $U_{\tau}$ values in BG data sets require corrections that may be estimated using the EKP near-wall profile as follows. Figure~10 of EKP shows that the near-wall profile in wall jets is described well by $U_{+}=z_{+}+C_{4}z_{+}^{4}$ where $C_{4}=-0.0004$ \citep{eriksson1998} represents maximum departure of the profile from linear variation. Now a Stanton tube is expected to respond to the dynamic pressure exerted by the average flow velocity in the gap between the test surface and the chamfered sharp leading edge of the steel blade. Therefore, one may compute the inner-scaled average velocity $U_{avg+}$ in a wall jet over the thickness of the blade using the above-mentioned wall-jet inner layer profile. Similarly, $U_{avg+}$ may be computed using the linear velocity profile which is typical of the channel-flow calibration of a Stanton tube. For the same value of $U_{avg}$, one may then compute the ratio of these two inner-scaled average velocities to obtain the ratio of actual $U_{\tau}$ value in the wall jet to that read by the Stanton tube as per its channel-flow calibration. For the four flows of BG (BG1-1 to BG1-4), these ratios respectively turn out to be $1.0172$, $1.0102$, $1.007$ and $1.005$. Therefore the $U_{\tau}$ values for BG data listed in table~\ref{tab:data} have been corrected by multiplying the original $U_{\tau}$ values by these enhancement factors.

The fourth data set belongs to the well-cited experiments of \cite{tailland1967} - henceforth TM. These are air-based experiments with three different nozzle Reynolds numbers ($\Rey_{j}=11000$, $18000$ and $25000$) as seen in table~\ref{tab:data}. Four representative hotwire mean velocity profiles have been digitized from figure~2 in TM. Nozzle is \emph{not} located in a wall perpendicular to the test surface and $\tau_{w}$ has been measured using velocity gradient in the near wall region. It is known that the $\tau_{w}$ values in TM are $20$ to $35\%$ lower than the consensus impact tube data \citep{launder1979} which is most likely due to the wall proximity effect of hotwires described above. In view of this, we have corrected skin friction values in TM data sets using an enhancement of $10\%$ in $U_{\tau}$ (i.e. $20\%$ in $\tau_{w}$).  

Finally, the fifth data set is taken from the recent DNS of \cite{naqavi2018} - henceforth NTT - having $\Rey_{j}=7500$. Nozzle is \emph{not} located in a wall perpendicular to the test surface because a co-flow velocity ($6\%$ of $U_{j}$) has been used as the inlet boundary condition above the nozzle as well as the outer boundary condition. The flow used in the present analysis corresponds to $x/b=35$ which is the farthest downstream station from the nozzle exit free from the domain end effects. Since the wall jet flow attains fully-developed state beyond $x/b=30$ as mentioned before, we have not used the data for $x/b<35$ from this DNS study. In addition to the mean velocity profile data at $x/b=35$, we have also used the Reynolds shear stress profiles to investigate layered structure based on mean momentum balance (\S~\ref{subsec:layers}).

\begin{figure}
\centerline{\includegraphics[width=0.8\textwidth]{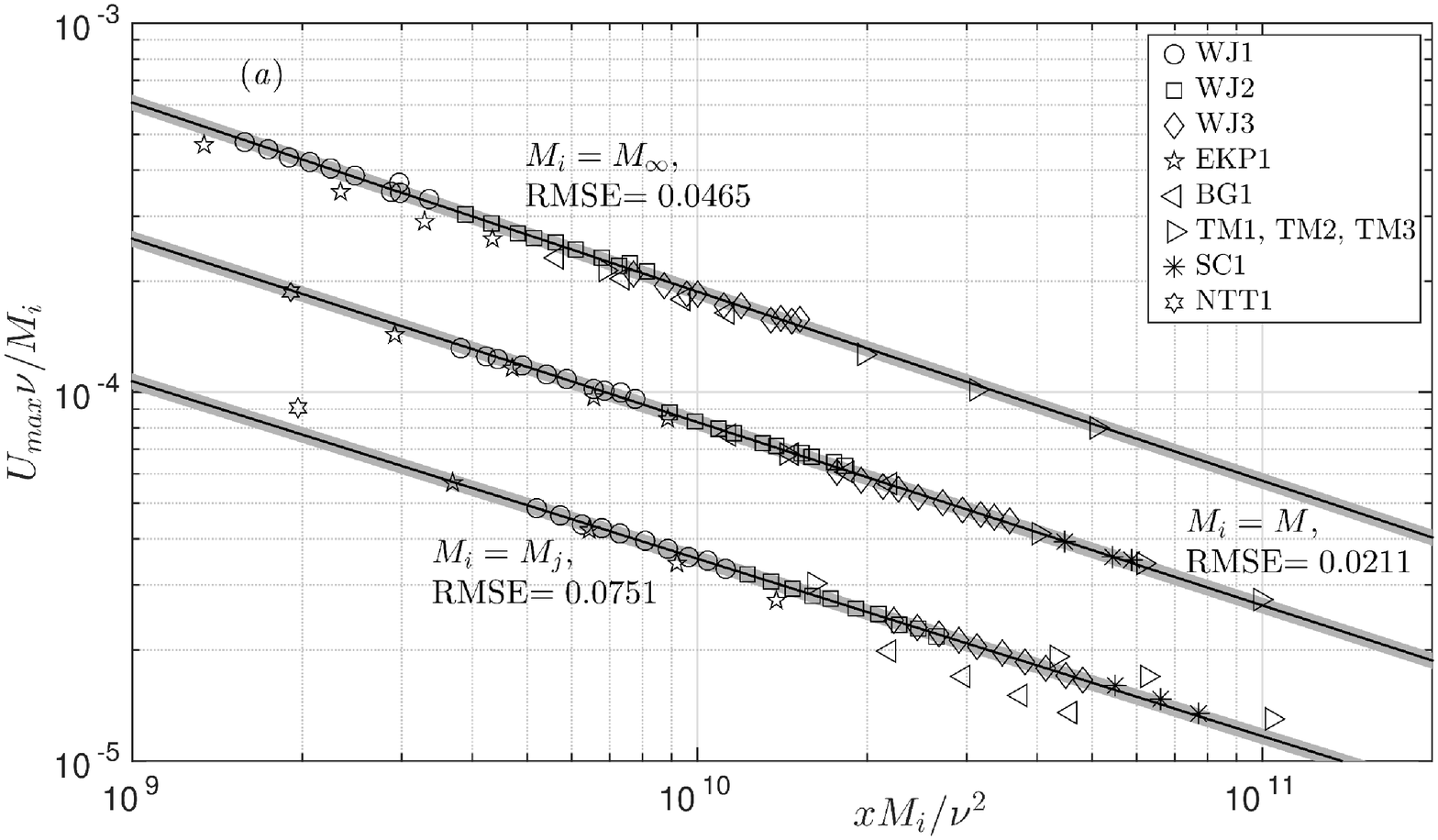}}   \centerline{\includegraphics[width=0.8\textwidth]{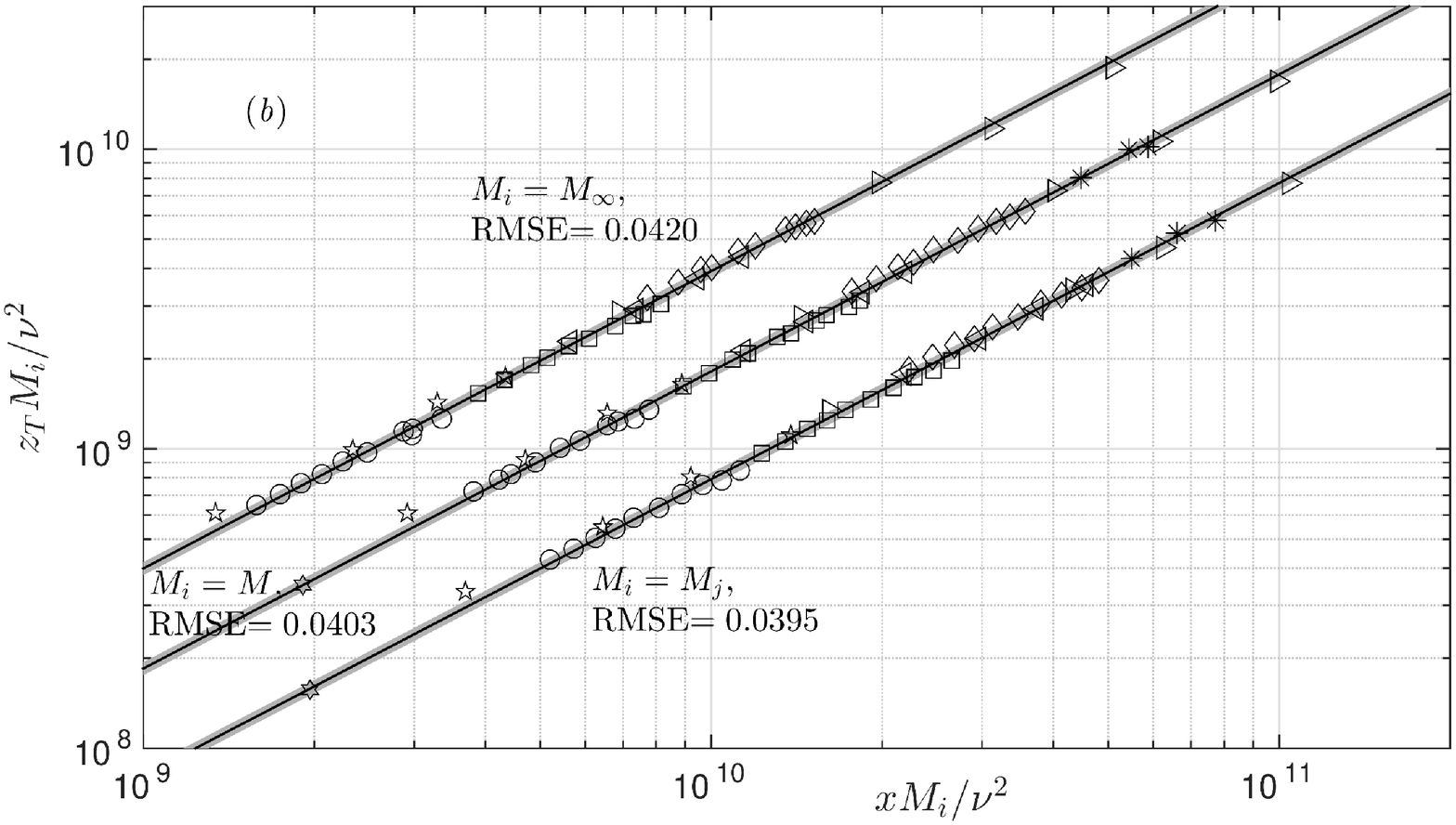}}   \centerline{\includegraphics[width=0.8\textwidth]{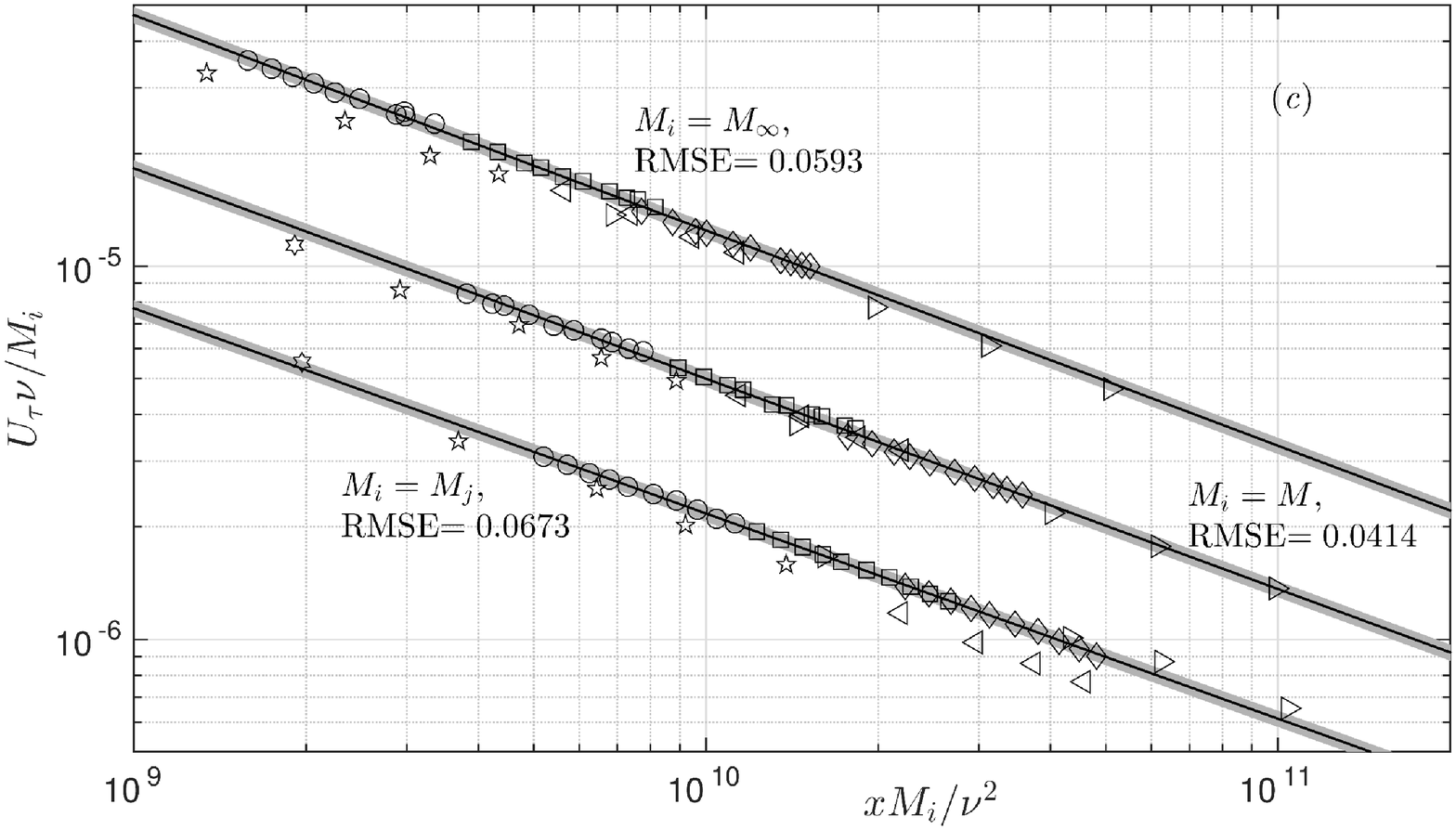}}
  \caption{Streamwise variations of (\textit{a}) $U_{max}$, (\textit{b}) $z_{T}$ and (\textit{c}) $U_{\tau}$ in the $M_{j}$-$\nu$ scaling of NYP, the presently proposed ``local" $M$-$\nu$ scaling and the $M_{\infty}$-$\nu$ scaling of \cite{gersten2015}. Data come from the experiments and simulations listed in table~\ref{tab:data}. Each solid line shows least-squares power law fitted to our data (WJ1, WJ2 and WJ3 in table~\ref{tab:data}). To avoid clutter in each plot, data points and fitted curve in categories $M_{i}=M$ and $M_{i}=M_{\infty}$ have been shifted upward using suitable arbitrary multiplying factors. Shading shows $\pm5\%$ band around each curve fit. For each scaling alternative, the RMS error of all data points with respect to the curve fit is also shown.}. 
\label{fig:MjnuandMnuscaling}
\end{figure}

There have been some recent water-based wall jet experiments \citep{tachie2002,rostamy2011,tang2015} that have used a facility similar to the EKP facility and covered similar range of Reynolds numbers. However, we have noticed a certain peculiar behaviour of this facility. Upon coming out from the nozzle, the kinematic momentum rate $M$ of the wall jet has been initially found to increase downstream up to $x/b\approx 50$ and reach a maximum value of $M/M_{j}\approx 1.25$ \citep[see][]{tachie2001}. Beyond this point, the momentum rate starts reducing in the downstream direction. Although there has been some explanation provided for this anomalous increase of momentum rate in this facility \citep{tachie2001}, no other wall jet studies have reported such a behaviour. Our contention is that this mean flow acceleration could be related to the secondary flow set up in the tank above the main wall jet flow. Since our scaling approach crucially hinges on the correctness of the local kinematic momentum rate $M$, these data sets had to be excluded from our analysis.   
 
\subsection{Scaling results}\label{subsec:scalingresults}

Figure~\ref{fig:MjnuandMnuscaling}(\textit{a}) shows the plot of $U_{max}$ as a function of $x$ for the data of table~\ref{tab:data}; note that $x$ is measured from the nozzle exit in each case and no adjustment for virtual origin has been made. Three scaling approaches using $M_{j}$, $M$ and $M_{\infty}$ (ICs, LCs and FCs of \textsection~\ref{subsec:ICs}, \textsection~\ref{subsec:LCs} and \textsection~\ref{subsec:FCs} respectively) are shown. Note that $M_{\infty}$ has been computed using the procedure given in \textsection~8 of \cite{gersten2015}. Also note that $xM_{\infty}/\nu^2$ is equivalent to $\Rey_{x}^{2}$ recommended by \cite{gersten2015}. Solid line in each case, shows the least-squares power law curve fit to our experimental data (WJ1, WJ2 and WJ3 in table~\ref{tab:data}) and the shading shows $\pm5\%$ band around the curve fit (and also for subsequent curve fits) to enable visual aid for assessing data collapse. To quantify the quality (goodness) of scaling with each approach, the normalized root-mean-squared error (RMSE) of all the data points is computed with respect to the corresponding curve fit. This RMSE value is mentioned alongside each curve fit; the lower the RMSE value the better is the scaling quality. It is clear that the a much better data collapse is seen with our \emph{local} scaling approach ($M$-$\nu$) i.e. with the LCs. Scaling using ICs ($M_{j}$-$\nu$) clearly shows larger data scatter. Scaling with FCs ($M_{\infty}$-$\nu$) also shows some data scatter but more importantly this approach has limitations since it does not apply to all flows due to nozzle located in a perpendicular wall \citep{gersten2015} in some cases such as the EKP data; these data are however included in the plot for completeness. Our approach (using LCs) does not suffer from this limitation and is applicable to all flows. Figures~\ref{fig:MjnuandMnuscaling}(\textit{b}) and \ref{fig:MjnuandMnuscaling}(\textit{c}) respectively show plots of $z_{T}$ and $U_{\tau}$ as functions of $x$. While $M_{j}$, $M$ and $M_{\infty}$ all give almost similar data collapse for $z_{T}$ (almost identical RMSE values in figure~\ref{fig:MjnuandMnuscaling}\textit{b}), $U_{\tau}$ clearly favours $M$-$\nu$ scaling (figure~\ref{fig:MjnuandMnuscaling}\textit{c}). Also, contrary to the expectation of NYP, our data indicate that $M/M_{j}$ is not a universal function of $xM_{j}/\nu^{2}$ (not shown). Further, $M_{\infty}/M_{j}$ is also not a universal constant \citep{gersten2015}. Therefore $M$ cannot be expressed in terms of either $M_{j}$ or $M_{\infty}$, and hence may be considered as an independent scaling parameter. 

In summary, the streamwise variations of $U_{max}$, $z_{T}$ and $U_{\tau}$ exhibit robust scaling behaviour in the \emph{local} $M$-$\nu$ framework irrespective of the facility, working fluid and whether the nozzle is located in a wall perpendicular to the test surface or not. Thus, basic condition for self-similarity i.e. lack of any imposed length or velocity scales, is met.   
    
\section{Scaling Mean Velocity Profiles}\label{sec:profilescaling}

Having demonstrated that the wall jet development is essentially governed by the local scaling parameters, we now proceed to the scaling of mean velocity profiles. Since ICs and FCs are not involved, the only possible complication that could now arise is the dependence of the velocity profile on local $\Rey_{\tau}$ (see \textsection~\ref{sec:intro}).

\subsection{Self-similar inner and outer regions}\label{subsec:innerouterscaling}

\begin{figure}
\centerline{\includegraphics[width=0.9\textwidth]{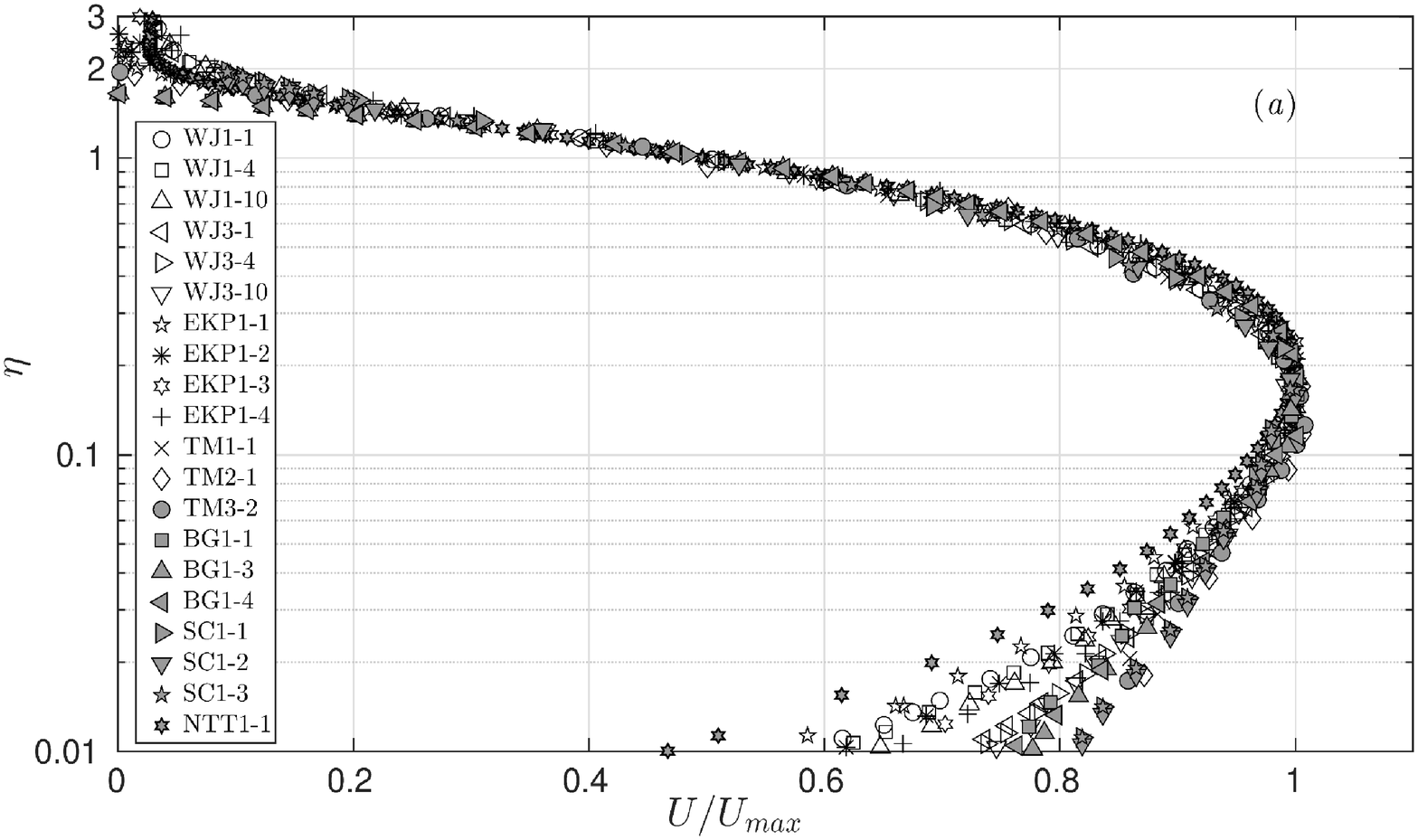}}
\centerline{\includegraphics[width=0.9\textwidth]{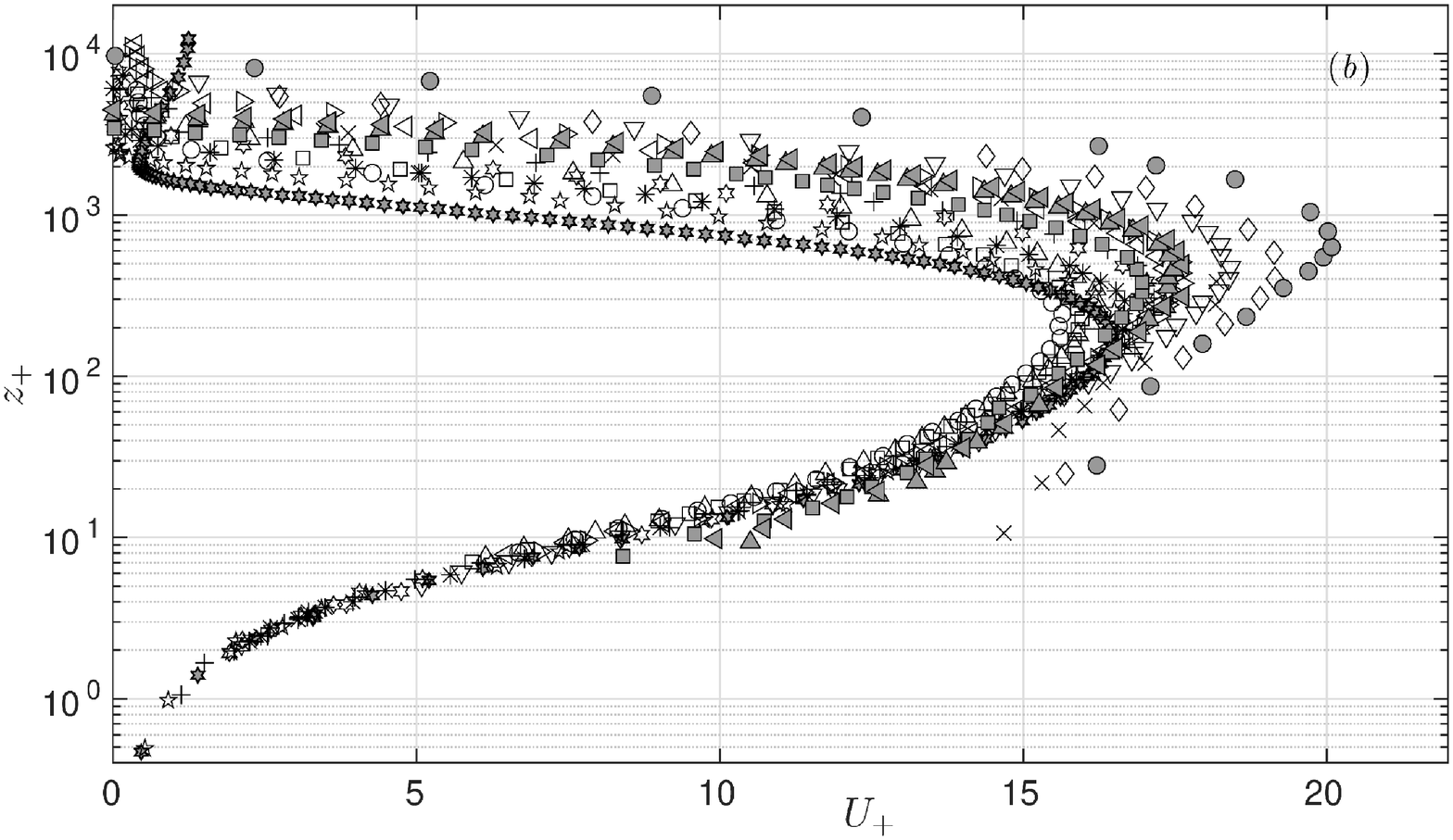}}
  \caption{Mean velocity profiles for the data sets listed in table~\ref{tab:data} in (\textit{a}) outer and (\textit{b}) inner coordinates. Plot (\textit{b}) does not contain SC data since $U_{\tau}$ was not measured in those experiments.}\label{fig:profilescaling}
\end{figure}

We first plot mean velocity profiles in classical outer or jet ($U/U_{max}$ versus $\eta$, figure~\ref{fig:profilescaling}\textit{a}) and inner or wall ($U_{+}$ versus $z_{+}$, figure~\ref{fig:profilescaling}\textit{b}) coordinates to look for any meaningful scaling; $\eta=z/z_{T}$, $U_{+}=U/U_{\tau}$ and $z_{+}=zU_{\tau}/\nu$. For the entire range of Reynolds numbers covered here, figure~\ref{fig:profilescaling}(\textit{a}) shows that all the data exhibit excellent \emph{scaling} (self-similarity) in the outer region including the velocity maximum and a small extent below it. The observation that the outer (jet) scaling holds even below the velocity maximum, suggests that the outer flow is in fact akin to a full-free jet rather than a half-free jet as contended by many other studies; this will be discussed later in \S~\ref{subsec:layers} in some detail. Note that the outer-scaled profiles show some scatter in the region close to the edge of the wall jet \citep[see figure~7 of][also]{george2000}. This could be due to the differences in the outer boundary condition such as reverse flow in the EKP setup or co-flow in the NTT data (see \S~\ref{subsec:data}) as well as larger uncertainties in hotwire measurements due to low velocities near the edge \citep{deo2008}. Figure~\ref{fig:profilescaling}(\textit{b}) shows the same data in the inner (wall) coordinates. BG and TM data clearly show the wall-proximity effect affecting the first few near-wall points in the hotwire profiles; the effect is very severe in TM profiles leading to spuriously high values of $U_{+}$. SC data are not plotted because $U_{\tau}$ has not been measured in their experiments. Notwithstanding these difficulties, remarkably robust scaling is evident for all the other data in the near-wall region (figure~\ref{fig:profilescaling}\textit{b}). Thus, the mean velocity data support classical inner and outer scalings (self-similarity) respectively of the form
\begin{eqnarray}
U_{+}&=&f\left(z_{+}\right),\label{eqn:inner}\\
U/U_{max}&=&g\left(\eta\right),\label{eqn:outer}
\end{eqnarray}
where $f$ and $g$ are \emph{universal} functions independent of $\Rey_{\tau}$ (at least to the lowest order). 

\subsection{Overlap layer analysis}\label{subsec:overlapanalysis}

\begin{figure}
\centerline{\includegraphics[width=0.9\textwidth]{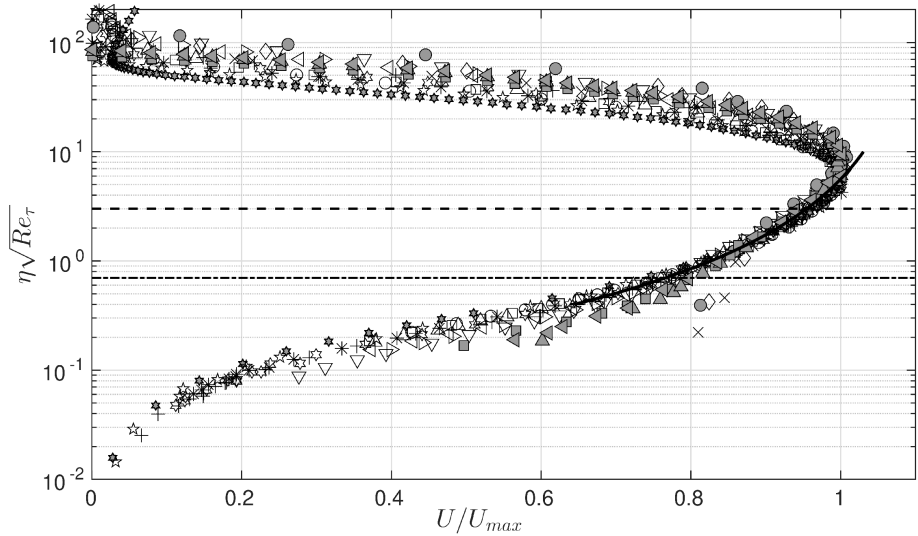}}
\caption{Mean velocity profiles for the data sets listed in table~\ref{tab:data} (except for the SC data) in the overlap layer scaling. Ordinate is the intermediate variable $\eta \sqrt{\Rey_{\tau}}$ or $z_{+}/\sqrt{\Rey_{\tau}}$. Solid line is the least-squares curve fit of (\ref{eqn:outerfinalpower2}) to all data between the dashed line and the dashed-dotted line ($0.7\leq\eta \sqrt{\Rey_{\tau}}\leq 3$). The curve is extended beyond these limits for visual aid. For symbols, refer to the legend of figure~\ref{fig:profilescaling}(\textit{a}).}
\label{fig:overlap}
\end{figure}
Figures~\ref{fig:profilescaling}(\textit{a}) and \ref{fig:profilescaling}(\textit{b}) indicate that the overlap of outer and inner scalings in wall jets is expected to occur below the velocity maximum. Matching $U$ and $\partial U/\partial z$ from the inner (\ref{eqn:inner}) and outer (\ref{eqn:outer}) descriptions in the overlap layer leads to 
\begin{equation}
\frac{\mathrm{d}\ln f}{\mathrm{d}\ln z_{+}}=\frac{\mathrm{d}\ln g}{\mathrm{d}\ln \eta}.\label{eqn:matching}
\end{equation}
Note that (\ref{eqn:matching}) is only superficially similar to (8.3) from \cite{george2000}; the main difference is that within the framework of George \emph{et al.}, $f$ and $g$ essentially depend on $\Rey_{\tau}$ whereas in (\ref{eqn:matching}) they are independent of $\Rey_{\tau}$ (due to \ref{eqn:inner} and \ref{eqn:outer}). Next, it is known that the solution to (\ref{eqn:matching}) is invariant under the transformation $z\rightarrow z+\alpha$ where $\alpha$ is an arbitrary shift in the origin for $z$ \citep[see][]{george2000}. Alternatively, we note that (\ref{eqn:matching}) also permits the transformation $U\rightarrow U+\beta$, where $\beta$ is an arbitrary shift in the origin for $U$. Physically either of these shifts accounts for a finite slip velocity at the wall since the overlap layer profile cannot satisfy the inner (no slip) boundary condition due to matching of the inner layer with the outer layer. In what follows, we replace $U$ by $\widetilde{U}=U+\beta$, so that $\widetilde{f}=\widetilde{U}/U_{\tau}=f+\beta_{+}$ and $\widetilde{g}=\widetilde{U}/U_{max}=g+\left(\beta/U_{max}\right)$. Since the left and right sides of (\ref{eqn:matching}) are purely functions of $z_{+}$ and $\eta$ respectively, each side must be equal to a \emph{universal} constant, say $A$. Integrating $\mathrm{d}\ln \widetilde{f}/\mathrm{d}\ln z_{+}=A$ leads to $\ln \widetilde{f}=A \ln z_{+} + B'$ in the overlap layer. Note that while $B'$ must be constant with respect to $z_{+}$ in the overlap layer, mathematically nothing precludes it from being a function of the local Reynolds number $\Rey_{\tau}$. Physically, this is closely related to the \emph{mesolayer} discussed in detail in \textsection\ref{subsec:layers}. Rewriting $B'=A\ln B + \ln D$ yields
\begin{equation}
\widetilde{f}=Dz_{+}^{A}B^{A},\label{eqn:innerpower}
\end{equation}
where $B\left(\Rey_{\tau}\right)$ and $D\left(\Rey_{\tau}\right)$ are unknown functions still to be determined. Similarly, $\mathrm{d}\ln \widetilde{g}/\mathrm{d}\ln \eta=A$ leads to
\begin{equation}
\widetilde{g}=E\eta^{A}C^{A},\label{eqn:outerpower}
\end{equation}
where $E\left(\Rey_{\tau}\right)$ and $C\left(\Rey_{\tau}\right)$ are functions unknown as yet. Clearly, (\ref{eqn:innerpower}) and (\ref{eqn:outerpower}) do \emph{not} scale the mean velocity profiles at different $\Rey_{\tau}$ in the overlap layer in classical inner and outer coordinates (figures~\ref{fig:profilescaling}\textit{a} and \ref{fig:profilescaling}\textit{b}). However, if one rearranges (\ref{eqn:innerpower}) and (\ref{eqn:outerpower}) as
\begin{eqnarray}
\left(\widetilde{f}/D\right)&=&\left(z_{+}B\right)^{A},\label{eqn:innermodipower}\\
\left(\widetilde{g}/E\right)&=&\left(\eta C\right)^{A},\label{eqn:outermodipower}
\end{eqnarray}
then $\widetilde{f}/D$ and $\widetilde{g}/E$ could become universal functions of $z_{+}B$ and $\eta C$ respectively. In order to proceed further with this contention, the functional forms of $B,C,D$ and $E$ are required to be determined.

Note that $z_{+}B$ and $\eta C$ are both akin to the so-called intermediate variable that is commonly used in the asymptotic analysis of multiscale problems \citep{kevorkian2013}. Physically, the role of function $B$($C$) is to ``rescale" $z_{+}$($\eta$) appropriately so that the overlap region remains in focus i.e. $z_{+}B$($\eta C$) remains of $\textit{O}(1)$ although $z_{+}\rightarrow\infty$($\eta\rightarrow 0$) as $\Rey_{\tau}\rightarrow\infty$. Therefore by the definition of an intermediate variable, $z_{+}B\sim\eta C$. Next, if one assumes power-law functional forms $B\sim \Rey_{\tau}^{m}$ and $C\sim \Rey_{\tau}^{n}$, then the condition $z_{+}B\sim\eta C$ leads to the constraint $1+m=n$ on the constants $m$ and $n$. In order to estimate the individual values of $m$ and $n$, we need additional information. Towards this, we note that since $\Rey_{\tau}$ assumes importance in the mean velocity overlap layer in wall jets (through functions $B$ and $C$ in \ref{eqn:innermodipower} and \ref{eqn:outermodipower} respectively), this layer could correspond to the \emph{mesolayer} that is studied extensively in the literature on mean momentum balance in wall-bounded flows \citep{george2000,wei2005}; it will be demonstrated later (\S~\ref{subsec:layers}) from DNS data that this indeed is the case. The momentum-balance mesolayer in canonical wall-bounded flows occurs around the location of maximum Reynolds shear stress where the Reynolds stress gradient term in the mean momentum equation looses its dynamical significance \citep{afzal1982,sreenivasan1997,wei2005} and this location scales as $z_{+}\sim \sqrt{\Rey_{\tau}}$. As we shall see later, DNS data show existence (very similar to other wall-bounded flows) of a momentum-balance mesolayer in wall jets as well. Therefore assuming that the mean-velocity overlap layer in the present analysis coincides with the momentum-balance mesolayer in wall jets, one has $z_{+}\sim \sqrt{\Rey_{\tau}}$ in the mean-velocity overlap layer. Since $z_{+}B\sim\textit{O}(1)$ in the mean-velocity overlap layer, it follows that $B\sim 1/\sqrt{\Rey_{\tau}}$ i.e. $m=-1/2$ and consequently the constraint $1+m=n$ yields $n=1/2$ i.e. $C\sim \sqrt{\Rey_{\tau}}$.

Since $z_{+}B\sim\eta C$ and $A$ is a universal constant, (\ref{eqn:innermodipower}) and (\ref{eqn:outermodipower}) imply $\left(\widetilde{f}/D\right)\sim\left(\widetilde{g}/E\right)$. Physically, functions $D$ and $E$ represent $\Rey_{\tau}$-dependent modulation of the inner and outer velocity scales respectively. For wall jets without an external stream, the jet momentum is expected to overwhelm the drag at the surface as $\Rey_{\tau}\rightarrow\infty$ \citep{gersten2015} with the asymptotic state being a half-free jet (see \textsection~\ref{subsec:layers}). Therefore the effect of the wall in the overlap region on the outer velocity scale ($\Rey_{\tau}$ dependence) may be expected to be very weak and enter only in the length scale. Therefore to the lowest order, it is reasonable to assume $E\left(\Rey_{\tau}\right)\approx\mathrm{constant}$ in the overlap layer. With this, $D\left(\Rey_{\tau}\right)\sim \widetilde{f}/\widetilde{g} \sim U_{max}/U_{\tau}$ and the velocity profile (\ref{eqn:innermodipower} and \ref{eqn:outermodipower}) in the overlap layer \emph{scales} according to either of the two equivalent relations
\begin{eqnarray}
\widetilde{f}/D&=&K_{i}\left(z_{+}/\sqrt{\Rey_{\tau}}\right)^{A},\label{eqn:innerfinalpower}\\
\widetilde{g}&=&K_{o}\left(\eta \sqrt{\Rey_{\tau}}\right)^{A},\label{eqn:outerfinalpower}
\end{eqnarray}
where $K_{i}$, $K_{o}$ and $A$ are universal constants. We shall now focus on (\ref{eqn:outerfinalpower}). Note that $\widetilde{g}=g+\left(\beta/U_{max}\right)$ where $\beta$ is constant for a given profile. Therefore $\beta/U_{max}$ could be a function of $\Rey_{\tau}$. However the same arguments that led to $E\left(\Rey_{\tau}\right)\approx\mathrm{constant}$, indicate that to the lowest order, one may expect $\beta/U_{max}\approx\mathrm{constant}$. With this, (\ref{eqn:outerfinalpower}) simplifies to
\begin{equation}
\frac{U}{U_{max}}=K_{o}\left(\eta \sqrt{\Rey_{\tau}}\right)^{A}-                                                                                                                                 \frac{\beta}{U_{max}},\label{eqn:outerfinalpower2}
\end{equation}
indicating that the mean velocity profiles in the overlap region \emph{scale} (collapse) in the coordinates $U/U_{max}$ versus $\eta \sqrt{\Rey_{\tau}}$.

\subsection{Overlap scaling in the experimental data}\label{subsec:overlapexptdata}

Figure~{\ref{fig:overlap}} shows all the data in table~\ref{tab:data} (except the SC data) plotted in coordinates $U/U_{max}$ versus $\eta \sqrt{\Rey_{\tau}}$ as suggested by (\ref{eqn:outerfinalpower2}). Indeed, all the data collapse remarkably well in the region where overlap is expected to occur. The collapse extends over the range $0.7\leq\eta \sqrt{\Rey_{\tau}}\leq 3$ i.e. $42\%$ of a decade in $\eta \sqrt{\Rey_{\tau}}$; the extent of this region is decided empirically by observation. Least-squares fit of (\ref{eqn:outerfinalpower2}) to all the data of figure~\ref{fig:overlap} over this range, yields the values for the universal constants as $K_{o}=-0.2879$, $A=-0.5346$ and $\beta/U_{max}=-1.1145$. More data from future high-Reynolds number experiments on wall jets would help accurate determination of these universal constants.

Note that the position and extent of the mean-velocity overlap are fixed in $\eta \sqrt{\Rey_{\tau}}$ or $z_{+}/\sqrt{\Rey_{\tau}}$ and not in $\eta$ or $z_{+}$. In fact, with increasing Reynolds number, the position of the overlap layer is expected to shift towards the wall and its extent is expected to reduce in the $\eta$ coordinate; the opposite is expected to be true for the $z_{+}$ coordinate. We shall come back to this point in the next section.   

\subsection{Layered structure}\label{subsec:layers}

\begin{figure}
\centerline{\includegraphics[width=0.9\textwidth]{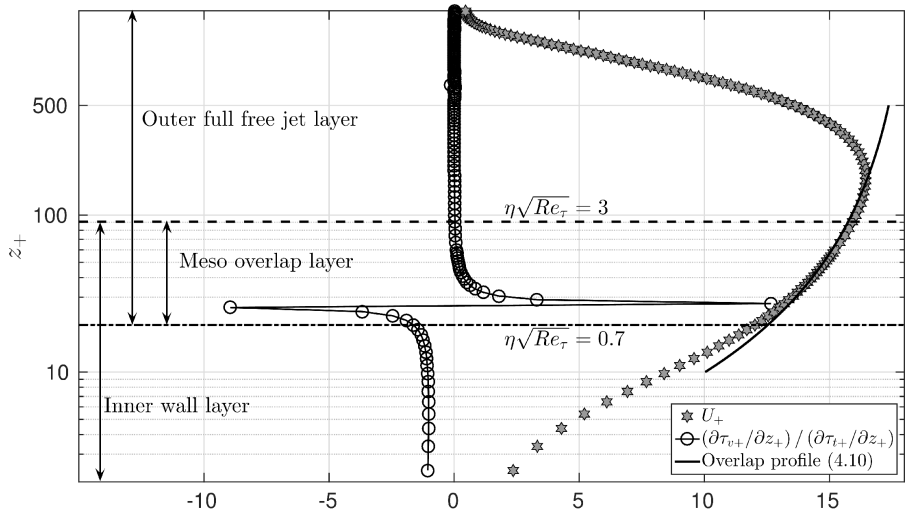}}
  \caption{Correspondence between the mean-velocity overlap layer and the momentum-balance mesolayer in wall jets. Inner-scaled profiles are shown for the mean velocity and the ratio of viscous stress gradient to the Reynolds shear stress gradient for the flow NTT1-1 (DNS data) of table~\ref{tab:data}. Dashed and dashed-dotted lines respectively denote the outer and inner limits $\eta\sqrt{\Rey_{\tau}}=3$ and $\eta \sqrt{\Rey_{\tau}}=0.7$ of the mean velocity overlap layer for $\Rey_{\tau}=865$. These limits have been fixed empirically in figure~\ref{fig:overlap}. Sharp flip in the stress gradient ratio is characteristic of the momentum-balance mesolayer.}\label{fig:DNS}
\end{figure}

In order to gain insight into the layered structure of wall jet flow, we plot in figure~\ref{fig:DNS} the profile of the ratio of viscous stress gradient ($\partial \tau_{v+}/\partial z_{+}=\partial^{2}U_{+}/\partial z_{+}^{2}$) to the Reynolds shear stress gradient ($\partial \tau_{t+}/\partial z_{+}=\partial (-\overline{u'_{+}w'_{+}})/\partial z_{+}$) for the DNS data (flow NTT1-1 in table~\ref{tab:data}) of \cite{naqavi2018}; $u'$ and $w'$ are velocity fluctuations in the streamwise and wall-normal directions respectively, and overbar denotes time average. Also plotted alongside are the mean velocity profile and the overlap layer profile (\ref{eqn:outerfinalpower2}). We observe that as far as the mean momentum balance is concerned, the layered structure of a wall jet is qualitatively very similar to that of channels and TBLs \citep{wei2005}. This suggests that the momentum balance approach does not distinguish between the jet or wake character of the outer layer in these flows, a fact that is not surprising given the nonlinear character of the Navi\'{e}r-Stokes equations and differences in boundary conditions in these flows. The innermost region is characterised by the balance of viscous and Reynolds shear stress gradients (ratio $\approx -1$) with advection being negligibly small. In the outer region, advection balances the Reynolds shear stress gradient with the viscous stress gradient being negligibly small (ratio $\approx 0$). The sharp flip in the ratio $(\partial \tau_{v+}/\partial z_{+})/(\partial \tau_{t+}/\partial z_{+})$ from large negative to large positive values indicates the mesolayer where the Reynolds shear stress reaches a maximum and viscous effects dominate the dynamics even at distances much further away from the viscous sublayer. The extent of the mean-velocity overlap layer for this flow is shown by horizontal dashed and dashed-dotted lines in figure~\ref{fig:DNS}. It is clear that the mean-velocity overlap layer, whose extent is decided empirically from the mean velocity data (figure~\ref{fig:overlap}), matches quite well with the momentum-balance mesolayer. This \emph{a posteriori} justifies the assumption $z_{+}\sim\sqrt{\Rey_{\tau}}$ for the mean-velocity overlap layer in wall jets and provides a dynamical basis for the matching arguments of \S~\ref{subsec:overlapanalysis}. Despite the fact that the DNS flow might still not have reached the fully-developed state (largest $x/b$ is $35$ only) and the flow Reynolds number is also relatively low ($\Rey_{\tau}=865$ only), the mean velocity profile in the overlap layer in figure~\ref{fig:DNS} is quite close to the curve of (\ref{eqn:outerfinalpower2}). 

Two important quantitative differences, however, exist when one compares wall jets with canonical channel and TBL flows. First, it may be noted that for channel and TBL flows, the location of the Reynolds shear stress maximum (or the location of the flip in the stress gradient ratio and hence the mesolayer) moves outwards from the wall in inner coordinates with increasing Reynolds number. For $\Rey_{\tau}\approx590$, the maximum is located at $z_{+}\approx45$ for channel flow DNS data \citep[see figure~1 of][]{wei2005}. Similarly for $\Rey_{\tau}\approx650$, the maximum is located at $z_{+}\approx55$ for TBL DNS data \citep[see figure~3 of][]{wei2005}. For the wall jet data NTT1-1 of table~\ref{tab:data} and figure~\ref{fig:DNS}, however, the Reynolds shear stress maximum occurs much closer to the wall i.e. below $z_{+}=30$ at $\Rey_{\tau}=865$. This observation is consistent with a thinner linear sublayer in wall jets compared to canonical flows (see \S~\ref{subsec:data}) and clearly shows that for the same Reynolds number, the influence of the outer full-jet flow penetrates deeper into the near-wall region in wall jets than the influence of the outer wake flow in the case of channels and TBLs. Second, the width of the mesolayer in channels and TBLs in $z_{+}$ coordinate is approximately equal to $\sqrt{\Rey_{\tau}}$ \citep{wei2005}. For wall jets, on the other hand, the width of the mean-velocity overlap layer is approximately $2.3\sqrt{\Rey_{\tau}}$ in $z_{+}$ coordinate (see figure~\ref{fig:overlap}). Thus the mean-velocity overlap layer in wall jets appears to be wider than the mesolayer in channels and TBLs.

\begin{figure}
\centerline{\includegraphics[width=0.9\textwidth]{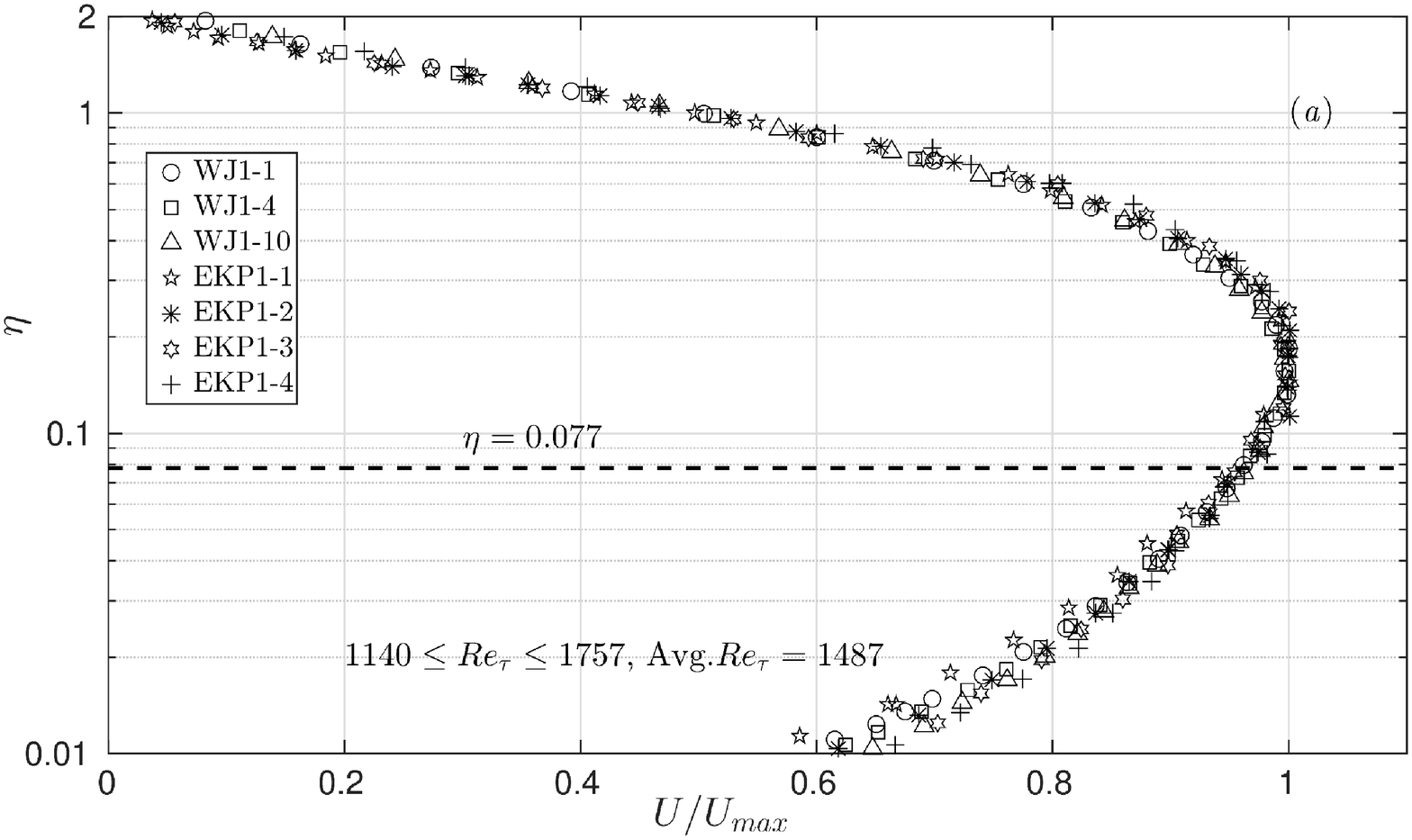}}
\centerline{\includegraphics[width=0.9\textwidth]{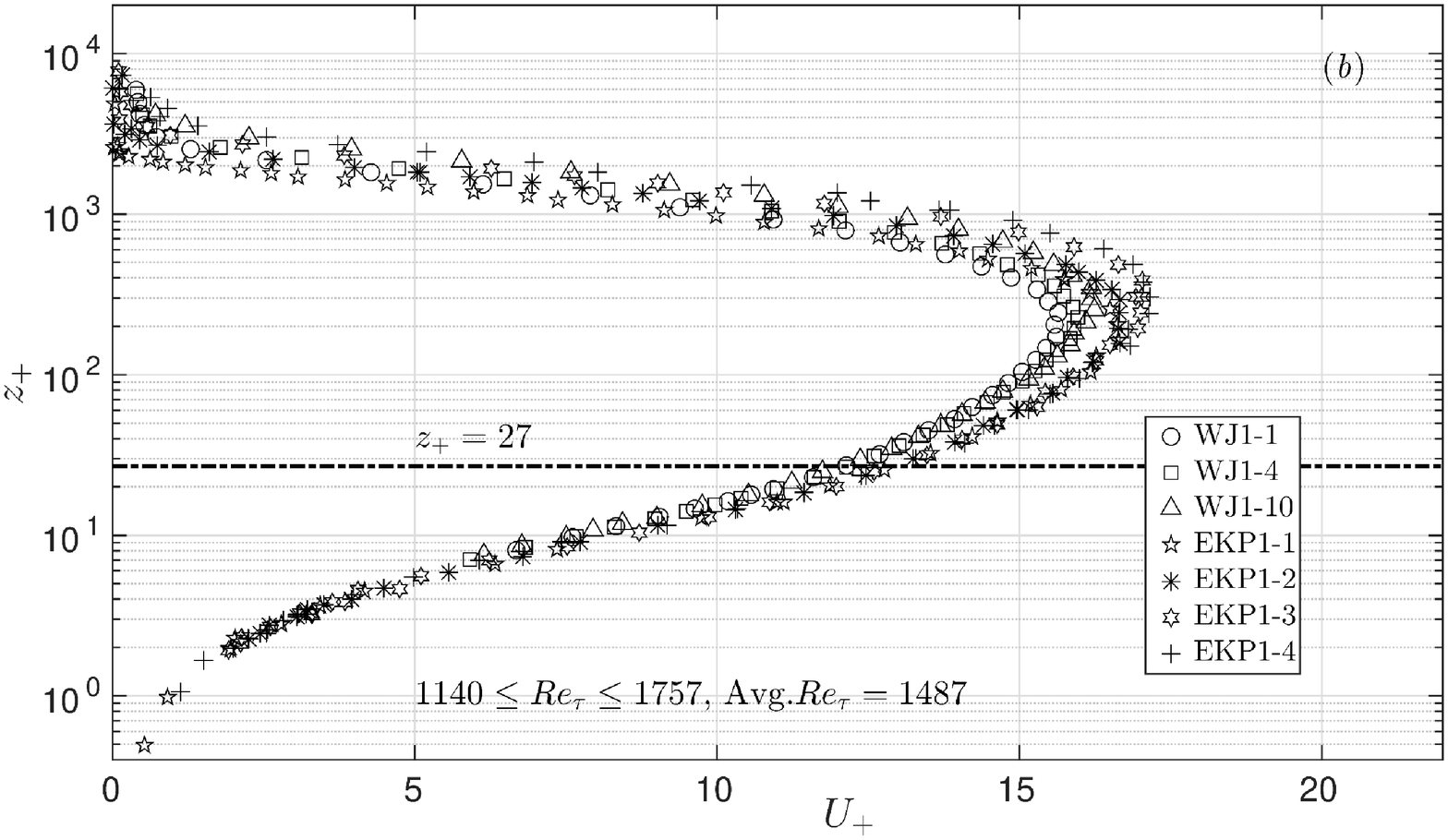}}
  \caption{(\textit{a}) Outer and (\textit{b}) inner scaling for low-Re wall jets of table~\ref{tab:data}. Dashed line in (\textit{a}) denotes the outer limit $\eta=0.077$ of the overlap layer computed from $\eta \sqrt{\Rey_{\tau}}=3$ for the average $\Rey_{\tau}=1487$. Dashed-dotted line in (\textit{b}) denotes the inner limit $z_{+}=27$ of the overlap layer computed from $\eta \sqrt{\Rey_{\tau}}=0.7$ for the same average $\Rey_{\tau}$.}\label{fig:lowRescaling}
\end{figure}

\begin{figure}
\centerline{\includegraphics[width=0.9\textwidth]{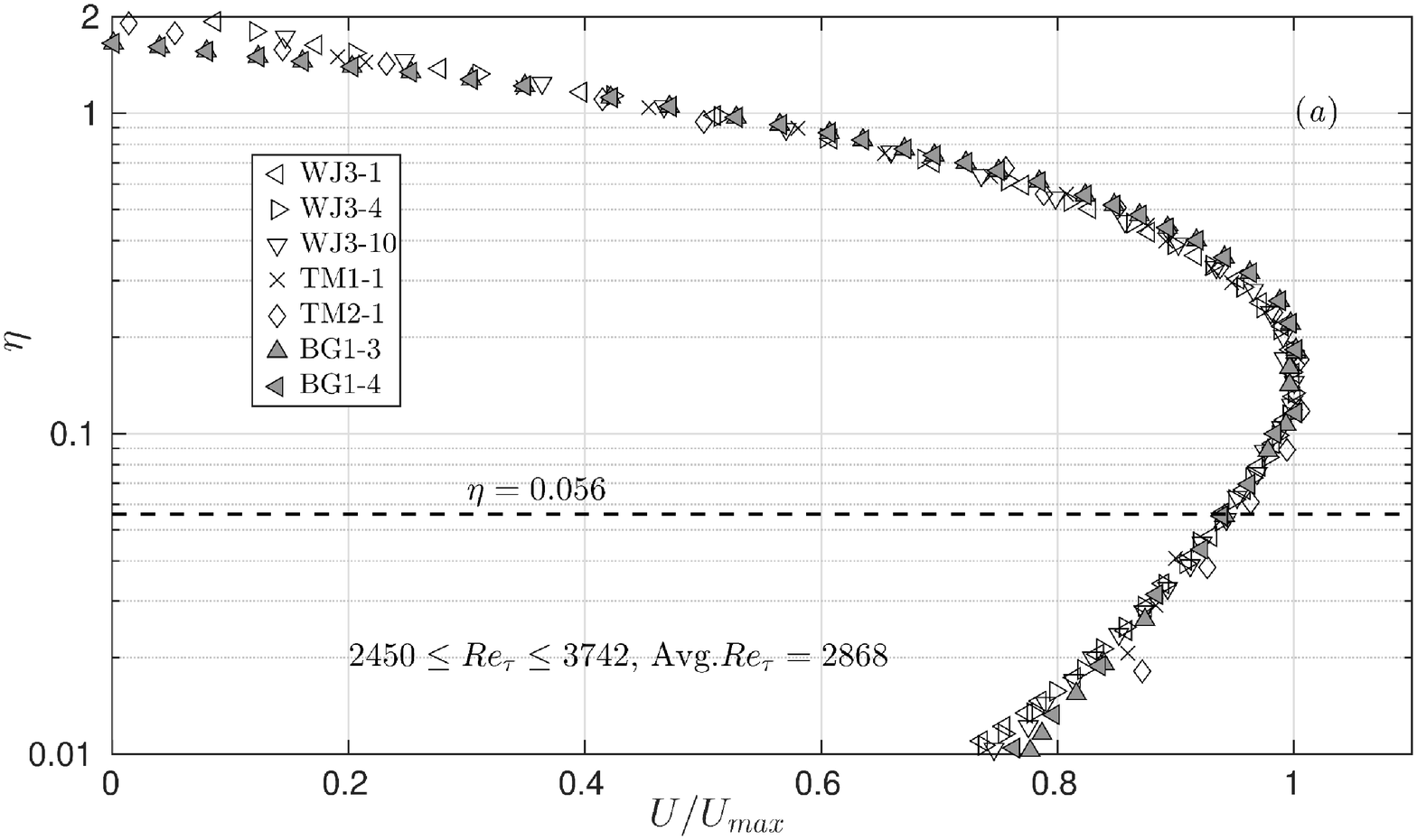}}
\centerline{\includegraphics[width=0.9\textwidth]{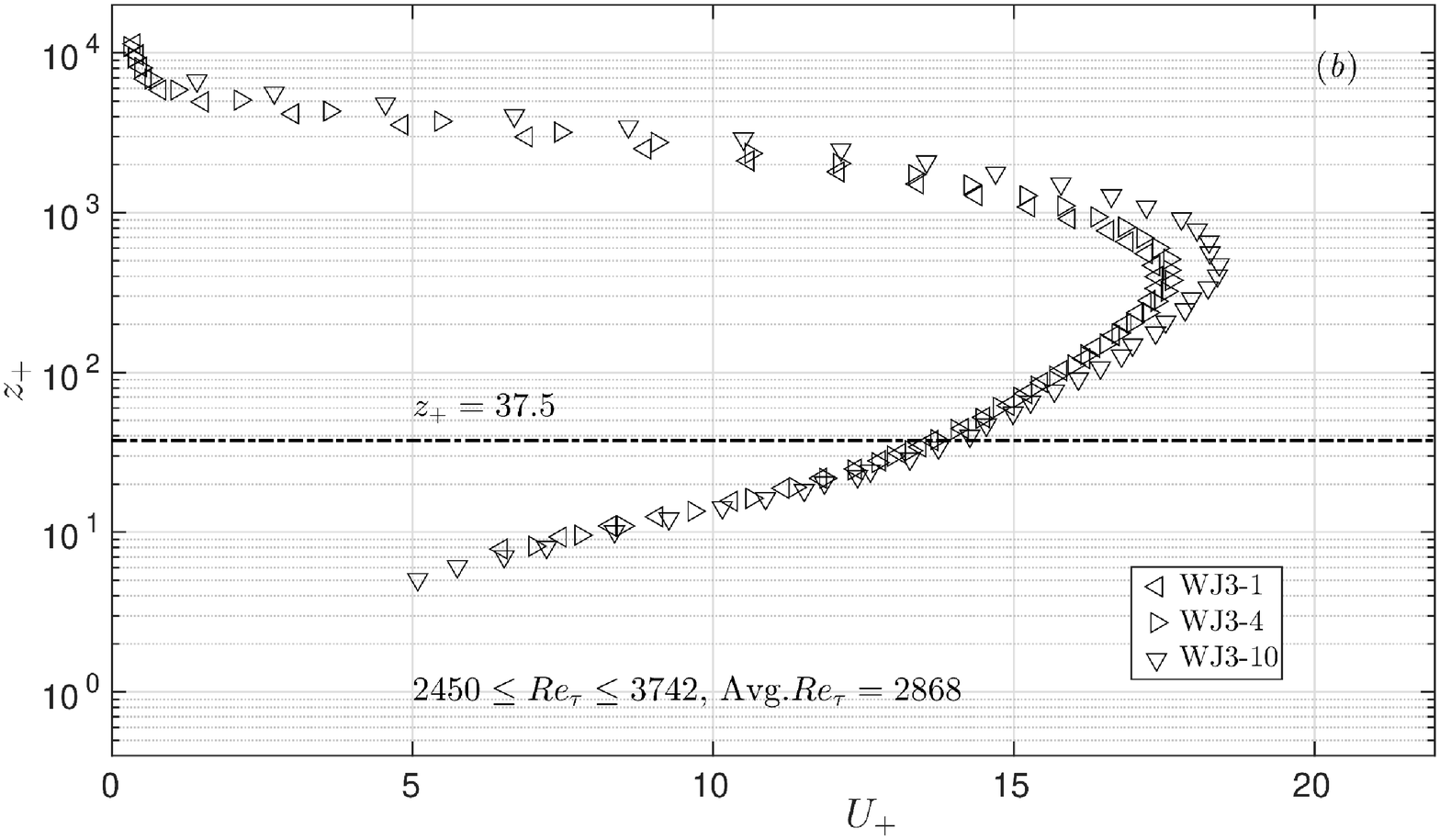}}
  \caption{(\textit{a}) Outer and (\textit{b}) inner scaling for high-Re wall jets of table~\ref{tab:data}. Dashed line in (\textit{a}) denotes the outer limit $\eta=0.056$ of the overlap layer computed from $\eta \sqrt{\Rey_{\tau}}=3$ for the average $\Rey_{\tau}=2868$. Dashed-dotted line in (\textit{b}) denotes the inner limit $z_{+}=37.5$ of the overlap layer computed from $\eta \sqrt{\Rey_{\tau}}=0.7$ for the same average $\Rey_{\tau}$.}\label{fig:highRescaling}
\end{figure}

A further question that now arises is whether there exists any inertial overlap layer between the inner and outer scaling regions. Usually the mesolayer overlap is considered to be the near-wall part (with $\Rey_{\tau}$-dependence) of the entire overlap layer and is followed by the inertial overlap (independent of $\Rey_{\tau}$) as one moves further outwards from the wall \citep{george2000}. To examine this, we plot the mean velocity profile data (table~\ref{tab:data}) by dividing it into two broad categories, low-$\Rey_{\tau}$ and high-$\Rey_{\tau}$. We plot the low-$\Rey_{\tau}$ data in figure~\ref{fig:lowRescaling} with the average $\Rey_{\tau}=1487$ and the range $1140\leq\Rey_{\tau}\leq1757$. Outer and inner limits ($\eta\sqrt{\Rey_{\tau}}=3$ and $\eta\sqrt{\Rey_{\tau}}=0.7$) of the mean-velocity overlap layer translate to $\eta=0.077$ and $z_{+}=27$ respectively for the average $\Rey_{\tau}=1487$. These are shown in figures~\ref{fig:lowRescaling}(\textit{a}) and~\ref{fig:lowRescaling}(\textit{b}) respectively by dashed and dashed-dotted lines. Figure~\ref{fig:highRescaling} shows the high-$\Rey_{\tau}$ data with average $\Rey_{\tau}=2868$ and the range $2450\leq\Rey_{\tau}\leq3742$ plotted in a fashion similar to figure~\ref{fig:lowRescaling}. Figures~\ref{fig:lowRescaling}(\textit{a}) and ~\ref{fig:highRescaling}(\textit{a}) clearly show that the outer scaling extends well below the velocity maximum supporting the present contention that the outer flow is akin to a full-free jet. The most striking observation from figures~\ref{fig:lowRescaling} and \ref{fig:highRescaling}, however, is that beyond the outer limit (in $\eta$ coordinate) of the mean-velocity overlap layer, data show remarkable collapse in outer scaling, and below the inner limit (in $z_{+}$ coordinate) of the overlap layer, data collapse quite well in the inner scaling. This implies that only the $\Rey_{\tau}$-dependent mean velocity overlap (mesolayer) of figure~\ref{fig:overlap} exists in wall jets and the inertial overlap layer does not appear to exist, at least for the range of Reynolds numbers considered here. Also it may be noted that with increase in Reynolds number, the outer edge of the overlap layer in figure~\ref{fig:highRescaling}(\textit{a}) shifts towards the wall in outer ($\eta$) coordinate as compared to figure~\ref{fig:lowRescaling}(\textit{a}). This indicates that the outer full-free jet flow in wall jets progressively dominates the inner wall flow with increasing Reynolds number. This is to be contrasted with the outer edge of the mean-velocity inertial sublayer (logarithmic overlap layer) being \emph{fixed} at $\eta\approx0.15$ in channels and TBLs \citep{wei2005,smits2011}. 

\begin{figure}
\centerline{\includegraphics[width=1.0\textwidth]{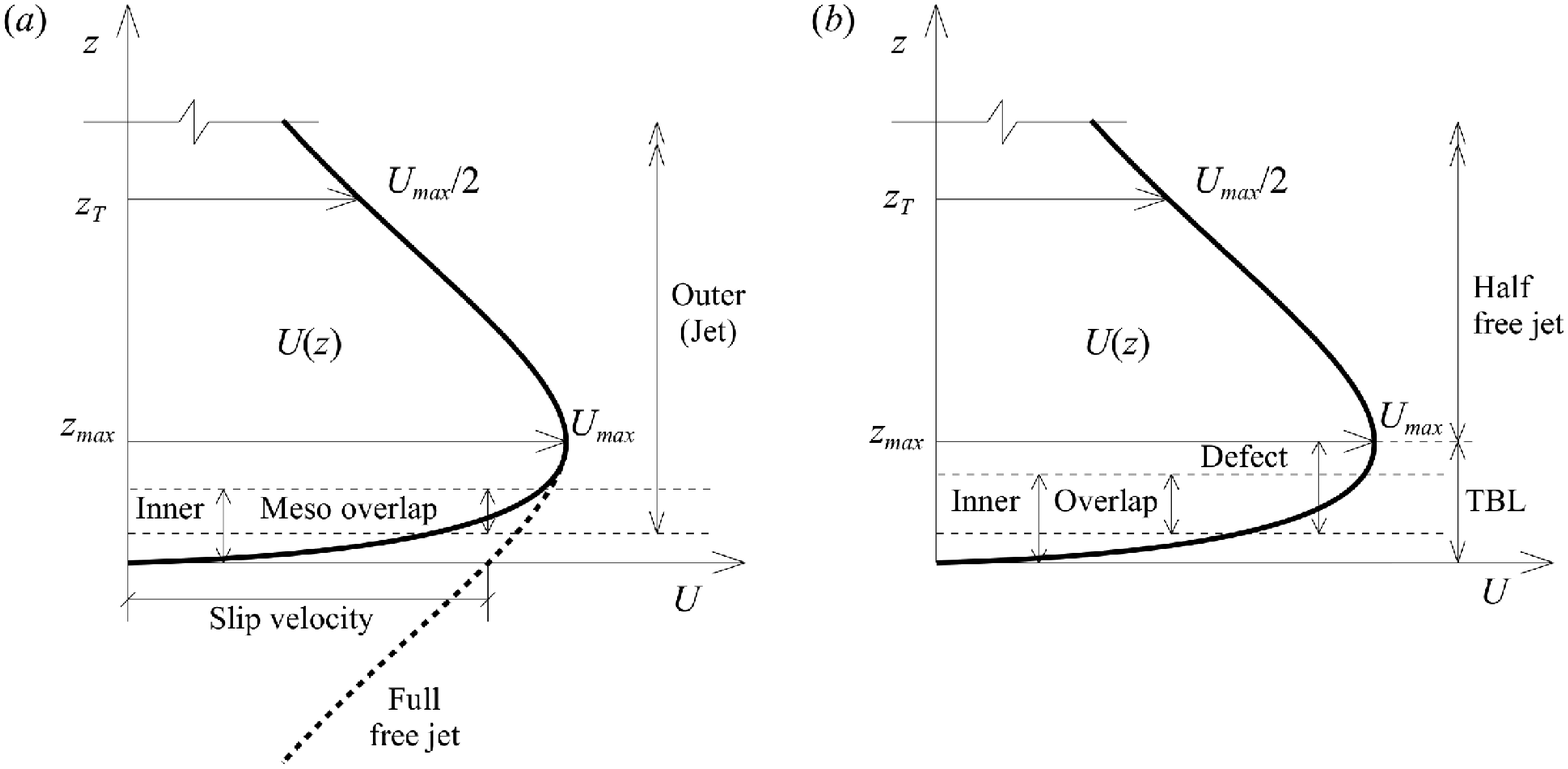}}
  \caption{(\textit{a}) Schematic of the layered structure of wall jet flow as per data and our analysis. Outer layer is a full-free jet centered on the velocity maximum having non-zero slip velocity at the wall. The part of the outer layer velocity profile shown by the thick dashed line gets modified due to overlap with the inner wall layer. The overlap layer is $\Rey_{\tau}$-dependent and coincides with the momentum-balance mesolayer. There is no inertial overlap between inner and outer scalings. (\textit{b}) Schematic of the prevalent proposal of the layered structure of wall jets from the literature \citep{afzal2005,gersten2015}. Part below $U_{max}$ is akin to a TBL, the part above is a half-free jet and these parts merge smoothly at the velocity maximum. Overlap occurs between the defect and inner regions of the TBL part of the flow.}\label{fig:layers}
\end{figure}

Thus, the layered structure of wall jets that emerges from the data and our analysis in the preceding may be summarized schematically as shown in figure~\ref{fig:layers}(\textit{a}). The outer flow is a self-similar turbulent full-free jet centered on the velocity maximum and described by the outer layer scaling (\ref{eqn:outer}). The outer jet layer does not satisfy the no-slip boundary condition at the wall as seen in figure~\ref{fig:layers}(\textit{a}). An inner or wall region, governed by (\ref{eqn:inner}), is therefore required to satisfy the no-slip condition at the wall and provide inner boundary condition for the outer flow. These inner and outer regions overlap in the form of an $\Rey_{\tau}$-dependent mean-velocity overlap layer that coincides with the momentum-balance mesolayer. There appears to be no inertial overlap between inner and outer scaling regions.

Note that the structure depicted in figure~\ref{fig:layers}(\textit{a}) supports the asymptotic half-free jet state proposed by \cite{afzal2005} and \cite{gersten2015} as will be discussed shortly. However in terms of the spirit and the details, the present proposal is distinct and minimal compared to the alternatives presented in other theoretical studies in the literature \citep{george2000,afzal2005,barenblatt2005,gersten2015}. The most prevalent alternative \citep{afzal2005,gersten2015} of these is the structure depicted in figure~\ref{fig:layers}(\textit{b}). Here the part of the flow below the velocity maximum $U_{max}$ behaves similar to a TBL with its own inner and defect scaling regions and a logarithmic (or power-law) overlap. The part above $U_{max}$ is a half-free jet and these parts patch up at the velocity maximum. The contention in this approach is that the asymptotic state far downstream would be a half free jet i.e. with increasing $\Rey_{\tau}$, the TBL part occupies increasing lesser fraction of the overall flow thickness. Note that the proposal of an asymptotic half-free jet state is quite plausible and consistent with our approach as well. Indeed, our data for WJ1, WJ2 and WJ3 experiments in table~\ref{tab:data} show that $U_{max}/U_{\tau}$ and $\Rey_{\tau}=z_{T}/\left(\nu/U_{\tau}\right)$ for each flow increase continually with $x/b$ implying more rapid decay of inner scales than the outer scales and this is consistent with the asymptotic half-free jet state. In fact, contrary to the expectation of \cite{gersten2015}, EKP data from tabel~\ref{tab:data} also show similar tendency towards a half-free jet state. However the contention that the entire flow below $U_{max}$ behaves like a TBL appears to be rather untenable \citep{bradshaw1963}. The occurrence of a region of counter-gradient momentum diffusion below $U_{max}$ in wall jets \citep{narasimha1990,ahlman2007,naqavi2018} is in stark contrast with \emph{always} down-the-gradient diffusion in the outer region of a TBL flow. A relatively recent LES study \citep{dejoan2005} shows that the anisotropy parameters, profiles of production-dissipation ratio etc. in wall jets differ radically from the corresponding turbulent channel flow results. Given the known similarity between channel flow and TBL structure, it is highly improbable that the part of wall jet flow just below $U_{max}$ has structure similar to the outer region of a TBL. The innermost region could still have similarities due to dominant role of the wall. Further detailed analyses of the experimental data of table~\ref{tab:data} following the approaches of \cite{afzal2005} and \cite{gersten2015} are presented in appendices~\ref{appA} and~\ref{appB} respectively. 

The layered structure of wall jets proposed by \cite{george2000} is summarized in figure~2 therein. While the structure we have proposed is very similar to their proposal, there are distinct differences such as $\Rey_{\tau}$-independent inner and outer scalings, absence of inertial overlap, arbitrary shift in $U$ instead of $z$ in the overlap analysis etc. The outer analysis of George \emph{et al.} rests on the contention that the Reynolds shear stress in the outer layer scales on $U_{\tau}$ rather than $U_{max}$. The experimental evidence (figure~10 of their paper) for this proposal is, however, very limited; there have been some recent studies that challenge this view \citep[see][]{ahlman2007,rostamy2011}. Theoretical argument in favour of this contention is presented in appendix~B of their paper where the outer equations reduce to a constant-stress layer in the overlap region. Again, the experimental evidence in wall jets for the occurrence of a constant-stress layer, or even a tendency towards it, is plainly absent (see figure~17 from their paper); this is in fact consistent with the absence of inertial overlap in wall jets. Our perception is that part of this confusion arises because the outer flow structure has not been identified correctly. With our structural model, the outer full-free jet flow does not lead to a constant-stress layer in the near-wall region due to non-negligible advection on account of large, finite slip velocity. More systematic experiments on wall jets at higher Reynolds numbers are required to shed more light on the correct scaling behaviour of the Reynolds shear stress.  

\cite{barenblatt2005} also have proposed layered structure for wall jets based on rather limited data of EKP experiments alone. In this proposal, the inner and outer layer length scales are respectively taken to be the heights of the half-velocity points ($U_{max}/2$) below and above the velocity maximum. Our view is that whilst the height of the upper half-velocity point could be justified as measure of the overall thickness of the flow \citep[as in turbulent free jets, see][]{deo2008}, the height of the lower half-velocity point seems rather adhoc and arbitrary choice without an immediately apparent physical significance (viscous length scale would be more appropriate in the near-wall region). The velocity scale $U_{max}$ essentially remains the same for both the layers. Furthermore, Barenblatt~\emph{et al.} do not discuss any overlap analysis.  

\section{Conclusions}\label{sec:conclusions}

We have addressed some key issues related to the scaling of mean velocity in fully-developed, two-dimensional, turbulent wall jets on flat surfaces. The conclusions may be summarized as follows. 
\begin{enumerate}
\item Streamwise variations of the velocity and length scales - namely $U_{max}$, $z_{T}$ and $U_{\tau}$ - follow ``local" scaling based on the kinematic momentum rate $M$ (per unit width) and kinematic viscosity $\nu$ (LCs) irrespective of the facility, working fluid and whether the nozzle is located in a wall perpendicular to the test surface or not. The flow development appears to be self-similar (layer-wise) without any imposed length and velocity scales i.e. independent of the inital conditions at the nozzle exit (ICs) as well as the asymptotic conditions far downstream (FCs). As is quite common in canonical developing flows such as TBLs, the inner and outer scales in a wall jet could still develop downstream at different rates. \label{i}
\item Experimental mean velocity data strongly support a two-layer self-similar structure of wall jets. The inner(wall) layer scales purely on $U_{\tau}$ and $\nu$ and the outer layer, which is akin to a full-free jet, scales purely on $U_{max}$ and $z_{T}$. These scalings are \emph{universal} i.e. independent of $\Rey_{\tau}$ (to the lowest order) as well as independent of the ICs and FCs.
\item Analysis shows that these inner and outer descriptions lead to an $\Rey_{\tau}$-dependent (non-universal) power-law velocity profile in the overlap layer of wall jets. Strong inner-outer interaction between the wall-flow structure and the highly-nonlinear full-free jet structure could lead to such an overlap.
\item Assumption of the mean-velocity overlap layer being coincident with the momentum-balance mesolayer enables description of the overlap layer in terms of an intermediate variable $\eta \sqrt{\Rey_{\tau}}$ (or $z_{+}/\sqrt{\Rey_{\tau}}$) that absorbs the $\Rey_{\tau}$-dependence; DNS data provide strong \emph{a posteriori} support for this assumption. With this, all the data (table~\ref{tab:data}) collapse remarkably well on to a single \emph{universal} power-law curve (\ref{eqn:outerfinalpower2}) for $42\%$ of a decade in the intermediate variable over the range $0.7<\eta \sqrt{\Rey_{\tau}}<3$. Since the overlap is fixed in $\eta \sqrt{\Rey_{\tau}}$, the position of the overlap layer shifts towards the wall and its extent reduces in the outer ($\eta$) coordinate; the opposite is true for the inner ($z_{+}$) coordinate.  
\item Analysis of the DNS data show that indeed the mean-velocity overlap layer, whose extent is determined empirically from the mean-velocity data, coincides well with the momentum-balance mesolayer. Further, it is observed that in wall jets the momentum-balance mesolayer occurs much closer to the wall (in $z_{+}$ coordinate) than in channels and TBLs at matched $\Rey_{\tau}$. This is consistent with the observation that the linear sublayer in wall jets extends only up to $z_{+}\approx3$ in contrast to $z_{+}\approx5$ in channels and TBLs. These observations show that for a given Reynolds number, the influence of the outer full-free jet flow structure penetrates deeper into the near-wall region as compared to the influence of the outer wake structure of channels and TBLs. This enables strong inner-outer interaction alluded to earlier. 
\item Selective plotting of experimental data for low- and high-$\Rey_{\tau}$ ranges shows that there is only $\Rey_{\tau}$-dependent overlap between the universal inner and outer mean-velocity scaling descriptions; no inertial overlap appears to exist in wall jets for the range of $\Rey_{\tau}$ values available till date. As opposed to channels and TBLs, the wallward extent of the outer scaling in wall jets increases with Reynolds number (outer scaling by itself is independent of $\Rey_{\tau}$ but its extent, specifically the lower end, depends on $\Rey_{\tau}$ due to the $\Rey_{\tau}$-dependent overlap) indicating that the outer full-free jet flow structure grows progressively stronger relative to the inner wall-flow structure. 
\end{enumerate}

\end{document}